\documentclass{article}
\usepackage{amsfonts}
\usepackage{makeidx}
\usepackage{latexsym,amsmath,amssymb,amscd}
\usepackage{makecell}
\usepackage{xcolor}
\usepackage{multirow}
\usepackage[all]{xy}

\allowdisplaybreaks
\newtheorem{theorem}{Theorem}

\newtheorem{proposition}[theorem]{Proposition}

\begin{document}

\title{Higher order first integrals of autonomous non-Riemannian dynamical systems}
\author{Antonios Mitsopoulos$^{1,a)}$, Michael Tsamparlis$^{2,3,b)}$ and Aniekan Ukpong$^{2,3,c)}$ \\
%EndAName
{\ \ }\\
$^{1}${\textit{Faculty of Physics, Department of
Astronomy-Astrophysics-Mechanics,}}\\
{\ \textit{University of Athens, Panepistemiopolis, Athens 157 83, Greece}}
\\
$^{2}${\textit{NITheCS, National Institute for Theoretical and Computational Sciences,}}\\
{\textit{Pietermaritzburg 3201, KwaZulu-Natal, South Africa}}\\
$^{3}${\textit{TCCMMP, Theoretical and Computational Condensed Matter and Materials Physics Group, }}\\
{\textit{School of Chemistry and Physics, University of KwaZulu-Natal,}}\\
{\textit{Pietermaritzburg 3201, KwaZulu-Natal, South Africa}}\\
\vspace{12pt} %EndAName
\\
$^{a)}$Author to whom correspondence should be addressed: antmits@phys.uoa.gr %EndAName
\\
$^{b)}$Email: mtsampa@phys.uoa.gr
\\
$^{c)}$Email: ukponga@ukzn.ac.za
\\
}
\date{}
\maketitle

\begin{abstract}
We consider autonomous holonomic dynamical systems defined by equations of the form $\ddot{q}^{a}=-\Gamma_{bc}^{a}(q) \dot{q}^{b}\dot{q}^{c}$ $-Q^{a}(q)$, where $\Gamma^{a}_{bc}(q)$ are the coefficients of a symmetric (possibly non-metrical) connection and $-Q^{a}(q)$ are the generalized forces. We prove a theorem which for these systems determines autonomous and time-dependent first integrals (FIs) of any order in a systematic way, using the `symmetries' of the geometry defined by the dynamical equations. We demonstrate the application of the theorem to compute linear, quadratic, and cubic FIs of various Riemannian and non-Riemannian dynamical systems.
\end{abstract}

\bigskip


\section{Introduction}

\label{sec.intro}

A first integral (FI) of a second order set of dynamical equations with generalized coordinates $q^{a}$ and generalized velocities $\dot{q}^{a}\equiv \frac{dq^{a}}{dt}$ is a function $I(t,q^{a},\dot{q}^{a})$ satisfying the condition $\frac{dI}{dt}=0$ along the dynamical equations. FIs are important because they can be used in order to reduce the order of the dynamical equations and, if they are `enough' in number \cite{Arnold 1989}, to find the solution of the system by quadrature (Liouville integrability).

The standard method to compute the FIs of Lagrangian systems is Noether's theorem. A different method is the direct method which requires only the dynamical equations and was originally introduced by Whittaker \cite{Whittaker, Katzin 1981, Thompson 1984, Horwood 2007, Post 2015, Mits 2021}. In the latter method, one assumes a functional form for the FI $I$ (e.g. a polynomial form in $\dot{q}^{a}$) and demands the condition $\frac{dI}{dt}=0$. Using the dynamical equations to remove the terms $\ddot{q}^{a}$ whenever they appear, the FI condition leads to a system of partial differential equations (PDEs) whose solution provides the FIs.

In this work, we apply the direct method to autonomous holonomic dynamical systems in a space with a symmetric connection $\Gamma_{bc}^{a}(q)$ (not necessarily Riemannian) which is read from the dynamical equations. We compute the resulting system of PDEs and solve it in terms of the `symmetries' of $\Gamma_{bc}^{a}(q)$. The result is stated as Theorem \ref{thm1} and provides a systematic method to determine polynomial FIs in velocities of any order, time-dependent and autonomous, for this type of dynamical systems. In the special case that the symmetric connection $\Gamma_{bc}^{a}(q)$ is the Riemannian connection defined by the kinetic metric (kinetic energy) $\gamma_{ab}(q)$ of the system, the computed FIs are directly related by means of the Inverse Noether Theorem \cite{Djukic, TsampMitsB} to gauged generalized (i.e. velocity-dependent) weak Noether symmetries. Finally, we apply Theorem \ref{thm1} in order to find new integrable and superintegrable systems which admit linear (LFIs), quadratic (QFIs), and cubic FIs (CFIs).

The structure of the paper has as follows. In section \ref{sec.conditions}, using the direct method, we derive the system of PDEs that must be satisfied by the coefficients of an $m$th-order FI of an autonomous (in general non-Riemannian) dynamical system. In section \ref{sec.theorem}, the `solution' of the system of PDEs is stated as Theorem \ref{thm1} (the proof of Theorem \ref{thm1} is given in the appendix). In section \ref{sec.lfis}, we apply Theorem \ref{thm1} for $m=1$ in order to consider the LFIs. It is shown that there are three independent types of LFIs and we determine their explicit formulae. In section \ref{sec.higherFIs}, we apply Theorem \ref{thm1} for $m>1$ and we find that there are three independent types of higher order FIs. In section \ref{sec.remove}, we discuss these independent FIs and provide a procedure by which symmetries of order smaller than the order of the FI can be removed. Using this procedure, we remove these lesser symmetries and we find the complete forms of the FIs for an even order $m=2\nu$ and an odd order $m=2\nu+1$. The results are collected in Propositions \ref{pro.thm1.4}, \ref{pro.thm1.5}, and \ref{pro.thm1.6}. In section \ref{sec.nonR}, we find a family of two-dimensional (2d) non-Riemannian autonomous dynamical systems and, by applying Theorem \ref{thm1}, we determine the LFIs. In sections \ref{sec.andro}, \ref{sec.appl2}, and \ref{sec.appl3}, we provide further applications of Theorem \ref{thm1} for QFIs and CFIs, by extending existing results in the literature. Finally, in section \ref{sec.conclusions}, we draw our conclusions.

\section{The conditions for $m$th-order FIs}

\label{sec.conditions}

We consider \emph{general} (i.e. Riemannian and non-Riemannian) autonomous dynamical systems of the form
\begin{equation}
\ddot{q}^{a}= -\Gamma^{a}_{bc}(q)\dot{q}^{b}\dot{q}^{c} -Q^{a}(q) \label{eq.tk1}
\end{equation}
where $q^{a}$ with $a=1,2,...,D$ are the generalized coordinates of the configuration space of the system, $D$ is the dimension of the configuration space, a dot over a letter indicates derivation with respect to (wrt) the parameter $t$ (time) along the trajectory $q^{a}(t)$, Einstein's summation convention is applied, $\Gamma^{a}_{bc}(q)$ are the coefficients of a general connection and $-Q^{a}(q)$ are the generalized forces. Since only the symmetric part $\Gamma^{a}_{(bc)}$ contributes to the dynamical equations --without loss of generality-- the quantities $\Gamma^{a}_{bc}(q)$ are assumed to be symmetric.

We look for $m$th-order FIs of the general form
\begin{equation}
I^{(m)}= \sum_{r=0}^{m} M_{i_{1}i_{2}...i_{r}}(t,q) \dot{q}^{i_{1}} \dot{q}^{i_{2}}...\dot{q}^{i_{r}}= M+ M_{i_{1}}\dot{q}^{i_{1}} +M_{i_{1}i_{2}} \dot{q}^{i_{1}} \dot{q}^{i_{2}} +...+M_{i_{1}i_{2}...i_{m}}\dot{q}^{i_{1}} \dot{q}^{i_{2}} ...\dot{q}^{i_{m}} \label{FI.5}
\end{equation}
where $M_{i_{1}...i_{r}}(t,q)$ with $r=0,1,...,m$ are totally symmetric $r$-rank tensors and the index $m \geq 1$ denotes the order of the FI. \emph{We note that when $r=0$, the quantities $M_{i_{1}...i_{r}}(t,q)$ reduce to the scalar $M(t,q)$.} For $m=1$, we have the LFIs; for $m=2$, the QFIs; and for $m=3$, the CFIs.

The FI condition
\begin{equation}
\frac{dI^{(m)}}{dt}=0  \label{DS1.10a}
\end{equation}%
along the dynamical equations (\ref{eq.tk1}) results in the following system of PDEs
\begin{equation}
M_{i_{1}i_{2}...i_{r},t} +M_{(i_{1}i_{2}...i_{r-1}|i_{r})} -(r+1)M_{i_{1}i_{2}...i_{r}i_{r+1}}Q^{i_{r+1}} =0, \enskip r=0, 1, 2, ..., m, m+1 \label{eq.veldep4} \\
\end{equation}
which is expanded as follows:
\begin{eqnarray}
M_{(i_{1}i_{2}...i_{m}|i_{m+1})} &=&0  \label{eq.veldep4.1} \\
M_{i_{1}i_{2}...i_{m},t}+M_{(i_{1}i_{2}...i_{m-1}|i_{m})} &=&0
\label{eq.veldep4.2} \\
M_{i_{1}i_{2}...i_{r},t}+M_{(i_{1}i_{2}...i_{r-1}|i_{r})} -(r+1)M_{i_{1}i_{2}...i_{r}i_{r+1}}Q^{i_{r+1}} &=&0, \enskip r=1,2,...,m-1, \enskip m>1 \label{eq.veldep4.3} \\
M_{,t}-M_{i_{1}}Q^{i_{1}} &=&0. \label{eq.veldep4.4}
\end{eqnarray}
The symbol $|$ denotes the covariant derivative wrt the symmetric connection $\Gamma^{a}_{bc}$, a comma indicates partial derivative wrt $q^{a}$ or $t$, round/square brackets indicate symmetrization/antisymmetrization of the enclosed indices, and indices enclosed between wavy lines are overlooked by symmetrization or antisymmetrization symbols.

We note that equation (\ref{eq.veldep4.1}) is derived from (\ref{eq.veldep4}) for $r=m+1$; equation (\ref{eq.veldep4.2}) from (\ref{eq.veldep4}) for $r=m$; and equation (\ref{eq.veldep4.4}) from (\ref{eq.veldep4}) for $r=0$.

Concerning the notation, we remark that:
\[
M_{i_{1}...i_{r-k}}(r=0)=
\begin{cases}
M, \enskip k=0 \\
0, \enskip k\geq 1
\end{cases}, \enskip
M_{i_{1}...i_{r}}(r>m)=0.
\]

Equations (\ref{eq.veldep4.1}) and (\ref{eq.veldep4.2}) are purely geometric equations, which are common to all systems of the form (\ref{eq.tk1}) that share the same symmetric connection. In particular, equation (\ref{eq.veldep4.1}) generalizes the concept of Killing tensors (KTs) to a non-metrical geometry with a symmetric connection $\Gamma^{a}_{bc}$. In this context, $M_{i_{1}i_{2}...i_{m}}$ is a \textbf{generalized $\mathbf{m}$th-order KT} for $\Gamma^{a}_{bc}$.

On the other hand, equations (\ref{eq.veldep4.3}) and (\ref{eq.veldep4.4}) are of a dynamical character, because they relate the geometric elements with the generalized forces $Q^{a}$ of the specific dynamical system.

Since the dynamical system (\ref{eq.tk1}) is autonomous, we should use the polynomial method described in \cite{Mits time} in order to solve the system of PDEs (\ref{eq.veldep4.1}) - (\ref{eq.veldep4.4}). According to this method, one assumes general polynomial expressions in the variable $t$ for the tensor quantities $M_{i_{1}...i_{r}}(t,q)$ with $r=1,2,...,m$ (see eq. (\ref{eq.aspm}) in the appendix), and replaces these expressions in the system of PDEs (\ref{eq.veldep4.1}) - (\ref{eq.veldep4.4}). Then, the scalar $M(t,q)$ is determined exactly, and the remaining PDEs reduce to polynomial equations in $t$ whose coefficients are functions of $q^{a}$. The `solution' of the latter system of PDEs is stated below as Theorem \ref{thm1}. A detailed proof is given in the appendix.

\section{Theorem for $m$th-order FIs of a general autonomous dynamical system}

\label{sec.theorem}

\begin{theorem} \label{thm1}
There are two types of $m$th-order FIs for the autonomous (in general non-Riemannian) dynamical system (\ref{eq.tk1}). These are the following:
\bigskip

\textbf{Integral 1.}
\begin{equation}
I^{(m)}_{n}= \sum_{r=1}^{m} \left( \sum^{n}_{N=0}L_{(N)i_{1}...i_{r}} t^{N}\right) \dot{q}^{i_{1}} ... \dot{q}^{i_{r}} +s_{0}\frac{t^{n+1}}{n+1} +\sum^{n>0}_{N=1} L_{(N-1)c}Q^{c} \frac{t^{N}}{N} +G(q) \label{eq.thm1.1}
\end{equation}
where $m\geq1$, $n\geq0$, $L_{(N)i_{1}...i_{m}}(q)$ with $N=0, 1, ..., n$ are $\mathbf{m}$\textbf{th-order generalized KTs} satisfying the condition
\begin{equation}
L_{(k)i_{1}...i_{m}} = -\frac{1}{k} L_{(k-1)(i_{1}...i_{m-1}|i_{m})}, \enskip k=1, 2, ..., n, \enskip n>0, \enskip m>1, \label{eq.thm1.2}
\end{equation}
the totally symmetric tensor $L_{(n)i_{1}...i_{m-1}}(q)$ with $m>1$ is an $\mathbf{(m-1)}$\textbf{th-order generalized KT}, the constant $s_{0}$ is defined by the condition
\begin{equation}
L_{(n)a}Q^{a} = s_{0} \label{eq.thm1.3}
\end{equation}
while the function $G(q)$ and the totally symmetric tensors $L_{(N)i_{1}...i_{r}}(q)$ satisfy the conditions:
\begin{eqnarray}
G_{,i_{1}} &=& 2L_{(0)i_{1}i_{2}}(m>1)Q^{i_{2}} -L_{(1)i_{1}}(n>0) \label{eq.thm1.4} \\
\left(L_{(k-1)c}Q^{c}\right)_{,i_{1}} &=& 2kL_{(k)i_{1}i_{2}}(m>1)Q^{i_{2}} -k(k+1)L_{(k+1)i_{1}}(k<n), \enskip k=1,2,...,n, \enskip n>0 \label{eq.thm1.5} \\
L_{(k)(i_{1}...i_{r-1}|i_{r})} &=& (r+1)L_{(k)i_{1}...i_{r}i_{r+1}} Q^{i_{r+1}} -(k+1) L_{(k+1)i_{1}...i_{r}}(k<n), \notag \\
&& k=0,1,..., n, \enskip r= 2,3,...,m-1, \enskip m>2. \label{eq.thm1.6}
\end{eqnarray}

\textbf{Integral 2.}
\begin{equation}
I^{(m)}_{e}= \frac{e^{\lambda t}}{\lambda} \left( \lambda \sum^{m}_{r=1} L_{i_{1}...i_{r}} \dot{q}^{i_{1}} ... \dot{q}^{i_{r}} +L_{c}Q^{c} \right) \label{eq.thm2.1}
\end{equation}
where $\lambda\neq0$, $L_{i_{1}...i_{m}}(q)$ is an $\mathbf{m}$\textbf{th-order generalized KT} satisfying the condition
\begin{equation}
L_{i_{1}...i_{m}}= -\frac{1}{\lambda} L_{(i_{1}...i_{m-1}|i_{m})}, \enskip m>1 \label{eq.thm2.2}
\end{equation}
and the totally symmetric tensors $L_{i_{1}...i_{r}}(q)$ satisfy the conditions:
\begin{eqnarray}
\left( L_{c}Q^{c} \right)_{,i_{1}}&=& 2\lambda L_{i_{1}i_{2}}(m>1) Q^{i_{2}} -\lambda^{2}L_{i_{1}} \label{eq.thm2.3} \\
L_{(i_{1}...i_{r-1}|i_{r})}&=& (r+1)L_{i_{1}...i_{r}i_{r+1}} Q^{i_{r+1}} -\lambda L_{i_{1}...i_{r}}, \enskip r=2,3,...,m-1, \enskip m>2. \label{eq.thm2.4}
\end{eqnarray}
\end{theorem}

Concerning the notation, the symbol $I^{(m)}_{n}$ means the $m$th-order FI (upper index) with degree of time-dependence $n$ (lower index), while $I^{(m)}_{e}$ indicates the $m$th-order FI with exponential time-dependence (lower index).

Using mathematical induction, one also proves the following recursion formulae.

\begin{proposition} \label{pro1}
The independent $m$th-order FIs $I_{n}^{(m)}$ and $I^{(m)}_{e}$ satisfy the following recursion formulae:\newline
a. $I_{n}^{(k)}<I_{n}^{(k+1)}$, that is, each $k$th-order FI $I_{n}^{(k)}$ is a subcase of the next $(k+1)$th-order FI $I_{n}^{(k+1)}$ with the same degree $n$ of time-dependence for all integers $k\geq1$. \newline
b. $I_{\ell}^{(m)}<I_{\ell+1}^{(m)}$, that is, the $m$th-order FI $I_{\ell}^{(m)}$ with time-dependence fixed by $\ell$ is a subcase of the $m$th-order FI $I_{\ell+1}^{(m)}$ with time-dependence $\ell+1$ for all integers $\ell\geq0$. \newline
c. $I_{e}^{(k)}<I_{e}^{(k+1)}$, that is, each $k$th-order FI $I_{e}^{(k)}$ is a subcase of the next $(k+1)$th-order FI $I_{e}^{(k+1)}$ for all integers $k\geq1$.
\end{proposition}

We note that Theorem \ref{thm1} for $m=2$ (QFIs) and a Riemannian connection reduces to Theorem 3 of \cite{TsampMitsB}.

In the case of a Riemannian connection, by means of the Inverse Noether Theorem \cite{Djukic, TsampMitsB}, the general $m$th-order FIs (\ref{FI.5}) are related to the generalized gauged weak Noether symmetry
\begin{equation}
\left( \xi=0, \enskip \eta_{i_{1}}= -\frac{\partial I^{(m)}}{\partial \dot{q}^{i_{1}}}, \enskip \phi_{a}, \enskip f =I^{(m)} -\frac{\partial I^{(m)}}{\partial \dot{q}^{i_{1}}}\dot{q}^{i_{1}} \right) \quad \text{such that $\phi_{a} \dot{q}^{a} +F^{a} \frac{\partial I^{(m)}}{\partial \dot{q}^{a}}= 0$} \label{eq.weak6}
\end{equation}
where $F^{a}(t,q,\dot{q})$ are the non-conservative generalized forces, $\phi^{a}(t,q,\dot{q})$ is an additional vector generator, $f(t,q,\dot{q})$ is the Noether function, $\mathbf{X}= \xi(t,q,\dot{q}) \partial_{t} + \eta^{a}(t,q,\dot{q}) \partial_{q^{a}}$ is the Lie generator, and the quantity
\[
\frac{\partial I^{(m)}}{\partial \dot{q}^{i_{1}}} = M_{i_{1}} + 2M_{i_{1}i_{2}}\dot{q}^{i_{2}} + 3M_{i_{1}i_{2}i_{3}} \dot{q}^{i_{2}} \dot{q}^{i_{3}} + ... + m M_{i_{1}i_{2}...i_{m}} \dot{q}^{i_{2}}...\dot{q}^{i_{m}} = \sum^{m-1}_{r=0} (r+1) M_{i_{1}i_{2}...i_{r+1}} \dot{q}^{i_{2}}...\dot{q}^{i_{r+1}}.
\]

\section{FIs of order $m=1$: LFIs}

\label{sec.lfis}

Applying Theorem \ref{thm1} for $m=1$ (LFIs), we find the following proposition for the LFIs.

\begin{proposition} \label{pro.thm1.1}
There are two types of LFIs for the autonomous (in general non-Riemannian) dynamical system (\ref{eq.tk1}) which are the following:
\bigskip

\textbf{LFI 1.}
\begin{equation}
I^{(1)}_{n}= \sum^{n}_{N=0}L_{(N)a} t^{N} \dot{q}^{a} +s_{0}\frac{t^{n+1}}{n+1} +\sum^{n>0}_{N=1} L_{(N-1)a}Q^{a} \frac{t^{N}}{N} +G(q) \label{eq.pro1.1}
\end{equation}
where $n\geq0$, $G(q)$ is an arbitrary smooth function, $L_{(N)a}(q)$ with $N=0, 1, ..., n$ are \textbf{generalized KVs} satisfying the conditions:
\begin{eqnarray}
L_{(1)a}(n>0)&=& -G_{,a}. \label{eq.pro1.2} \\
\left(L_{(k-1)b}Q^{b}\right)_{,a}&=& -k(k+1)L_{(k+1)a}(k<n), \enskip k=1,2,...,n, \enskip n>0, \label{eq.pro1.3}
\end{eqnarray}
and the constant $s_{0}$ is defined by the condition
\begin{equation}
L_{(n)a}Q^{a} = s_{0}. \label{eq.pro1.4}
\end{equation}

\textbf{LFI 2.}
\begin{equation}
I^{(1)}_{e}= \frac{e^{\lambda t}}{\lambda} \left( \lambda L_{a} \dot{q}^{a} +L_{a}Q^{a} \right) \label{eq.pro1.5}
\end{equation}
where $\lambda\neq0$ and $L_{a}(q)$ is a \textbf{generalized KV} satisfying the condition
\begin{equation}
\left( L_{b}Q^{b} \right)_{,a}= -\lambda^{2}L_{a}. \label{eq.pro1.6}
\end{equation}
\end{proposition}

The LFI (\ref{eq.pro1.1}) consists of two independent LFIs: a) A LFI for the even vectors $L_{(2N)a}$, and b) another one for the odd vectors $L_{(2N+1)a}$ and the scalar $G(q)$. This results directly from the fact that condition (\ref{eq.pro1.2}) involves only the odd vector $L_{(1)a}$ and the scalar $G(q)$, while conditions (\ref{eq.pro1.3}) are separated in two sets; one set involving only the even vectors $L_{(2N)a}$ (i.e. for the odd values $k=1,3,5,...$), and another one involving only the odd vectors $L_{(2N+1)a}$ (i.e. for the even values $k=2,4,6,...$). We note that the above holds for either an odd or an even time-dependence $n$. In the following proposition, we give the explicit formulae of these two independent LFIs.

\begin{proposition}
\label{pro.thm1.2} The LFI $I^{(1)}_{n}$ given in (\ref{eq.pro1.1}) consists of the following two independent LFIs:
\bigskip

\textbf{LFI 1.1.}
\begin{equation}
I^{(1,1)}_{\ell}= \sum_{N=1}^{\ell} t^{2N-1} L_{(2N-1)a} \dot{q}^{a} +s_{1}\frac{t^{2\ell}}{2\ell} +\sum_{N=1}^{\ell-1\geq1} L_{(2N-1)a}Q^{a} \frac{t^{2N}}{2N} +G(q) \label{eq.pro1.10}
\end{equation}
where $\ell>0$, $L_{(2N-1)a}(q)$ with $N=1,2,...,\ell$ are \textbf{generalized KVs} satisfying the condition
\begin{equation}
\left(L_{(2k-1)b}Q^{b}\right)_{,a}= -2k(2k+1)L_{(2k+1)a}, \enskip k=1,2,...,\ell-1, \enskip \ell>1, \label{eq.pro1.11}
\end{equation}
$G(q)$ is an arbitrary smooth function such that $L_{(1)a}= -G_{,a}$ is a \textbf{gradient generalized KV}, and the constant $s_{1}$ is defined by the condition
\begin{equation}
s_{1}= L_{(2\ell-1)a}Q^{a}. \label{eq.pro1.12}
\end{equation}

\textbf{LFI 1.2.}
\begin{equation}
I^{(1,2)}_{\ell}= \sum_{N=0}^{\ell} t^{2N}L_{(2N)a}\dot{q}^{a} +s_{0}\frac{t^{2\ell+1}}{2\ell+1} +\sum_{N=0}^{\ell-1\geq0} L_{(2N)a}Q^{a} \frac{t^{2N+1}}{2N+1} \label{eq.pro1.7}
\end{equation}
where $\ell \geq0$, $L_{(2N)a}(q)$ with $N=0,1,...,\ell$ are \textbf{generalized KVs} satisfying the condition
\begin{equation}
\left(L_{(2k)b}Q^{b}\right)_{,a}= -2(2k+1)(k+1)L_{(2k+2)a}, \enskip k=0,1,...,\ell-1, \enskip \ell>0, \label{eq.pro1.8}
\end{equation}
and the constant $s_{0}$ is defined by the condition
\begin{equation}
L_{(2\ell)a}Q^{a}= s_{0}. \label{eq.pro1.9}
\end{equation}

Therefore, the general autonomous dynamical system (\ref{eq.tk1}) admits three independent LFIs which are given by the formulae (\ref{eq.pro1.5}), (\ref{eq.pro1.10}), and (\ref{eq.pro1.7}).
\end{proposition}

The notation $I^{(1,\alpha)}_{\ell}$, where $\alpha=1,2$, indicates one of the two types of independent LFIs of the type $I^{(1)}_{n}$ with time-dependence fixed by the lower index $\ell$.

We note that condition (\ref{eq.pro1.11}) is derived from (\ref{eq.pro1.3}) for the even values $k=2,4,6,..., 2\ell$ if we set $n=2\ell$ and rename the index $k$ as $2k$; while condition (\ref{eq.pro1.8}) is derived from (\ref{eq.pro1.3}) for the odd values $k=1,3,5,..., 2\ell-1$ if we set $n=2\ell$ and rename the index $k$ as $2k+1$.

\section{FIs of order $m>1$}

\label{sec.higherFIs}

Applying Theorem \ref{thm1} for $m>1$, we find the following proposition.

\begin{proposition} \label{pro.thm1.3}
There are two types of $m$th-order FIs with $m>1$ for the autonomous (in general non-Riemannian) dynamical system (\ref{eq.tk1}) which are the following:
\bigskip

\textbf{Integral 1.}
\begin{equation}
I^{(m>1)}_{n}= \sum_{r=1}^{m} \left( \sum^{n}_{N=0}L_{(N)i_{1}...i_{r}} t^{N}\right) \dot{q}^{i_{1}} ... \dot{q}^{i_{r}} +s_{0}\frac{t^{n+1}}{n+1} +\sum^{n>0}_{N=1} L_{(N-1)c}Q^{c} \frac{t^{N}}{N} +G(q) \label{eq.pro2.1}
\end{equation}
where $m>1$, $n\geq0$, $L_{(N)i_{1}...i_{m}}(q)$ with $N=0, 1, ..., n$ are $\mathbf{m}$\textbf{th-order generalized KTs} satisfying the condition
\begin{equation}
L_{(k)i_{1}...i_{m}} = -\frac{1}{k} L_{(k-1)(i_{1}...i_{m-1}|i_{m})}, \enskip k=1, 2, ..., n, \enskip n>0, \label{eq.pro2.2}
\end{equation}
the totally symmetric tensor $L_{(n)i_{1}...i_{m-1}}(q)$ is an $\mathbf{(m-1)}$\textbf{th-order generalized KT}, the constant $s_{0}$ is defined by the condition
\begin{equation}
L_{(n)a}Q^{a} = s_{0} \label{eq.pro2.3}
\end{equation}
while the function $G(q)$ and the remaining totally symmetric tensors $L_{(N)i_{1}...i_{r}}(q)$ satisfy the conditions:
\begin{eqnarray}
G_{,i_{1}} &=& 2L_{(0)i_{1}i_{2}}Q^{i_{2}} -L_{(1)i_{1}}(n>0) \label{eq.pro2.4} \\
\left(L_{(k-1)c}Q^{c}\right)_{,i_{1}} &=& 2kL_{(k)i_{1}i_{2}}Q^{i_{2}} -k(k+1)L_{(k+1)i_{1}}(k<n), \enskip k=1,2,...,n, \enskip n>0 \label{eq.pro2.5} \\
L_{(k)(i_{1}...i_{r-1}|i_{r})} &=& (r+1)L_{(k)i_{1}...i_{r}i_{r+1}} Q^{i_{r+1}} -(k+1) L_{(k+1)i_{1}...i_{r}}(k<n), \notag \\
&& k=0,1,..., n, \enskip r= 2,3,...,m-1, \enskip m>2. \label{eq.pro2.6}
\end{eqnarray}

\textbf{Integral 2.}
\begin{equation}
I^{(m>1)}_{e}= \frac{e^{\lambda t}}{\lambda} \left( \lambda \sum^{m}_{r=1} L_{i_{1}...i_{r}} \dot{q}^{i_{1}} ... \dot{q}^{i_{r}} +L_{c}Q^{c} \right) \label{eq.pro2.7}
\end{equation}
where $\lambda\neq0$, $L_{i_{1}...i_{m}}(q)$ is an $\mathbf{m}$\textbf{th-order generalized KT} satisfying the condition
\begin{equation}
L_{i_{1}...i_{m}}= -\frac{1}{\lambda} L_{(i_{1}...i_{m-1}|i_{m})} \label{eq.pro2.8}
\end{equation}
and the remaining totally symmetric tensors $L_{i_{1}...i_{r}}(q)$ with $r=1,2,...,m-1$ satisfy the conditions:
\begin{eqnarray}
\left( L_{c}Q^{c} \right)_{,i_{1}}&=& 2\lambda L_{i_{1}i_{2}} Q^{i_{2}} -\lambda^{2}L_{i_{1}} \label{eq.pro2.9} \\
L_{(i_{1}...i_{r-1}|i_{r})}&=& (r+1)L_{i_{1}...i_{r}i_{r+1}} Q^{i_{r+1}} -\lambda L_{i_{1}...i_{r}}, \enskip r=2,3,...,m-1, \enskip m>2. \label{eq.pro2.10}
\end{eqnarray}
\end{proposition}

From Proposition \ref{pro.thm1.3}, we observe that the conditions (\ref{eq.pro2.2}) - (\ref{eq.pro2.6}) of the $m$th-order FI (\ref{eq.pro2.1}) are divided in two classes, according to if the tensor quantities $L_{(N)i_{1}...i_{r}}(q)$ are of even or odd order, and the associated index $N$ is even or odd. Therefore, the FI (\ref{eq.pro2.1}) consists of two independent FIs: \newline
a) One FI, say $I^{(m,1)}_{\ell}$, which contains tensor quantities $L_{(2N-1)i_{1}...i_{r}}(q)$ of odd order and $L_{(2N)i_{1}...i_{r}}(q)$ of even order. \newline
b) Another FI, say $I^{(m,2)}_{\ell}$, which contains tensor quantities $L_{(2N-1)i_{1}...i_{r}}(q)$ of even order and $L_{(2N)i_{1}...i_{r}}(q)$ of odd order.

We note that the order $r$ of the involved totally symmetric tensors $L_{(N)i_{1}...i_{r}}(q)$ takes values from $1$ to $m>1$, where $m$ is the order of the considered FI (\ref{eq.pro2.1}).

The independent FIs $I^{(m,1)}_{\ell}$ and $I^{(m,2)}_{\ell}$ are given by the following explicit formulae:
\bigskip

a.
\begin{eqnarray}
I^{(m>1,1)}_{\ell} &=& \sum_{r=1,\text{odd}}^{m} \sum_{N=1}^{\ell>0} t^{2N-1} L_{(2N-1)i_{1}...i_{r}} \dot{q}^{i_{1}} ... \dot{q}^{i_{r}} +\sum_{r=1,\text{even}}^{m} \sum_{N=0}^{\ell} t^{2N} L_{(2N)i_{1}...i_{r}} \dot{q}^{i_{1}} ... \dot{q}^{i_{r}} + \notag \\
&& +\sum_{N=1}^{\ell>0}L_{(2N-1)c}Q^{c} \frac{t^{2N}}{2N} +G(q) \label{eq.pro3.1}
\end{eqnarray}
where the time-dependence $\ell \geq0$ and the involved quantities satisfy the conditions:
\begin{eqnarray}
G_{,i_{1}}&=& 2L_{(0)i_{1}i_{2}}Q^{i_{2}} -L_{(1)i_{1}}(\ell>0) \label{eq.pro3.2} \\
\left(L_{(2N-1)c}Q^{c} \right)_{,i_{1}}&=& 4NL_{(2N)i_{1}i_{2}} Q^{i_{2}} -2N(2N+1)L_{(2N+1)i_{1}}(N<\ell), \enskip N=1,2,...,\ell, \enskip \ell>0 \label{eq.pro3.3} \\
L_{(2N-1)(i_{1}...i_{r-1}|i_{r})} &=& (r+1) L_{(2N-1)i_{1}...i_{r}i_{r+1}}Q^{i_{r+1}} -2NL_{(2N)i_{1}...i_{r}}, \notag \\
&& N=1,2,...,\ell, \enskip r=2,4,6,..., \enskip \ell>0 \label{eq.pro3.4} \\
L_{(2N)(i_{1}...i_{r-1}|i_{r})} &=& (r+1) L_{(2N)i_{1}...i_{r}i_{r+1}}Q^{i_{r+1}} -(2N+1)L_{(2N+1)i_{1}...i_{r}}(N<\ell), \notag \\
&& N=0,1,...,\ell, \enskip r=3,5,7,... ~. \label{eq.pro3.5}
\end{eqnarray}
In conditions (\ref{eq.pro3.4}) and (\ref{eq.pro3.5}), it holds that $r\leq m-1$ where $m>2$.

b.
\begin{eqnarray}
I^{(m>1,2)}_{\ell} &=& \sum_{r=1,\text{odd}}^{m} \sum_{N=0}^{\ell} t^{2N} L_{(2N)i_{1}...i_{r}} \dot{q}^{i_{1}} ... \dot{q}^{i_{r}} +\sum_{r=1,\text{even}}^{m} \sum_{N=1}^{\ell>0} t^{2N-1} L_{(2N-1)i_{1}...i_{r}} \dot{q}^{i_{1}} ... \dot{q}^{i_{r}} + \notag \\
&& +s_{0}\frac{t^{2\ell+1}}{2\ell+1} +\sum_{N=0}^{\ell-1\geq0} L_{(2N)c}Q^{c} \frac{t^{2N+1}}{2N+1} \label{eq.pro4.1}
\end{eqnarray}
where the time-dependence $\ell \geq0$ and the involved quantities satisfy the conditions:
\begin{eqnarray}
L_{(2\ell)a}Q^{a}&=&s_{0} \label{eq.pro4.0} \\
\left( L_{(2N-2)c}Q^{c} \right)_{,i_{1}} &=& 2(2N-1) L_{(2N-1)i_{1}i_{2}} Q^{i_{2}} -2N(2N-1)L_{(2N)i_{1}}, \enskip N=1,2,...,\ell, \enskip \ell>0 \label{eq.pro4.2} \\
L_{(2N)(i_{1}...i_{r-1}|i_{r})} &=& (r+1) L_{(2N)i_{1}...i_{r}i_{r+1}}Q^{i_{r+1}} -(2N+1)L_{(2N+1)i_{1}...i_{r}}(N<\ell), \notag \\
&& N=0,1,...,\ell, \enskip r=2,4,6,..., \label{eq.pro4.3} \\
L_{(2N-1)(i_{1}...i_{r-1}|i_{r})} &=& (r+1) L_{(2N-1)i_{1}...i_{r}i_{r+1}}Q^{i_{r+1}} -2NL_{(2N)i_{1}...i_{r}}, \notag \\
&& N=1,2,...,\ell, \enskip r=3,5,7,..., \enskip \ell>0. \label{eq.pro4.4}
\end{eqnarray}
In conditions (\ref{eq.pro4.3}) and (\ref{eq.pro4.4}), it holds that $r\leq m-1$ where $m>2$.

For the FIs (\ref{eq.pro3.1}) and (\ref{eq.pro4.1}), we have also that $L_{(2\ell)i_{1}...i_{m-1}}(q)$ is an \textbf{$\mathbf{(m-1)}$th-order generalized KT}, and $L_{(N)i_{1}...i_{m}}(q)$ with $N=0, 1, ..., 2\ell$ are \textbf{$\mathbf{m}$th-order generalized KTs} such that:
\begin{eqnarray}
L_{(2N)i_{1}...i_{m}}&=& -\frac{1}{2N} L_{(2N-1)(i_{1}...i_{m-1}|i_{m})}, \enskip N=1,2,...,\ell, \enskip \ell>0 \label{eq.pro0.1} \\
L_{(2N-1)i_{1}...i_{m}}&=& -\frac{1}{2N-1} L_{(2N-2)(i_{1}...i_{m-1}|i_{m})}, \enskip N=1,2,...,\ell, \enskip \ell>0. \label{eq.pro0.2}
\end{eqnarray}

We note that the sum $\sum_{r=1,\text{odd}}^{m}$ is over the odd values of $r$, while the sum $\sum_{r=1,\text{even}}^{m}$ is over the even values of $r$.

\section{How to remove the $(m-1)$th-order geometric symmetries from the two independent FIs $I^{(m>1,1)}_{\ell}$ and $I^{(m>1,2)}_{\ell}$: The complete forms}

\label{sec.remove}

As we have seen, the two independent FIs (\ref{eq.pro3.1}) and (\ref{eq.pro4.1}) include geometric symmetries of order smaller from the order $m>1$ of the FIs. These symmetries are described by the $(m-1)$th-order generalized KT $L_{(2\ell)i_{1}...i_{m-1}}(q)$. We observe that if $m=2\nu$ (even) where $\nu>0$, the considered $(m-1)$th-order generalized KT appears in the FI (\ref{eq.pro4.1}); while in the case that $m=2\nu+1$ (odd), appears in the FI (\ref{eq.pro3.1}). Moreover, for an even order $m$, the condition (\ref{eq.pro0.1}) accompanies the FI (\ref{eq.pro3.1}) and the condition (\ref{eq.pro0.2}) accompanies the FI (\ref{eq.pro4.1}); while for an odd order $m$, the condition (\ref{eq.pro0.1}) accompanies the FI (\ref{eq.pro4.1}) and the condition (\ref{eq.pro0.2}) accompanies the FI (\ref{eq.pro3.1}).

The $(m-1)$th-order generalized KT $L_{(2\ell)i_{1}...i_{m-1}}(q)$ can be removed from the expressions (\ref{eq.pro3.1}) and (\ref{eq.pro4.1}) by applying the following procedure:
\bigskip

1) \underline{For an even order $m=2\nu$ where $\nu>0$.}

In this case, the $(2\nu-1)$th-order generalized KT $L_{(2\ell)i_{1}...i_{2\nu-1}}(q)$ appears in the independent FI (\ref{eq.pro4.1}). We can remove this $(2\nu-1)$th-order symmetry by introducing a sequence of an even rank totally symmetric tensors of a higher time-dependence, that is, we add in the FI (\ref{eq.pro4.1}) terms of the form
\begin{equation}
t^{2\ell+1} L_{(2\ell+1)i_{1}...i_{r}} \dot{q}^{i_{1}} ... \dot{q}^{i_{r}}, \enskip r=2,4,...,2\nu, \enskip \nu>0 \label{eq.prf1.1}
\end{equation}
such that $L_{(2\ell+1)i_{1}...i_{2\nu}}$ is a $(2\nu)$th-order generalized KT given by the relation
\begin{equation}
L_{(2\ell+1)i_{1}...i_{2\nu}}= -\frac{1}{2\ell+1} L_{(2\ell)(i_{1}...i_{2\nu-1}|i_{2\nu})}. \label{eq.prf1.2}
\end{equation}
Then, the condition (\ref{eq.pro4.0}) must be generalized as
\begin{equation}
\left( L_{(2\ell)c}Q^{c} \right)_{,i_{1}}= 2(2\ell+1) L_{(2\ell+1)i_{1}i_{2}} Q^{i_{2}}; \label{eq.prf1.3}
\end{equation}
from the condition (\ref{eq.pro4.3}), the restriction $N<\ell$  must be removed because now quantities of the form $L_{(2N+1)i_{1}...i_{r}}$ do exist for $N=\ell$; and we must add the condition
\begin{equation}
L_{(2\ell+1)(i_{1}...i_{r-1}|i_{r})} = (r+1) L_{(2\ell+1)i_{1}...i_{r}i_{r+1}}Q^{i_{r+1}}, \enskip r=3,5,7,...,2\nu-1 \enskip \nu>1. \label{eq.prf1.4}
\end{equation}

Observe that the condition (\ref{eq.prf1.2}) removes the original $(2\nu-1)$th-order symmetry by allowing the $(2\nu-1)$th-order generalized KT $L_{(2\ell)i_{1}...i_{2\nu-1}}(q)$ to be a totally symmetric tensor of the same rank, which is not necessarily a KT. Then, the case of the generalized KT is derived as a subcase. This shows that symmetries of an order lesser than the order of the FI can always be absorbed into a higher order term.

2) \underline{For an odd order $m=2\nu+1$ where $\nu>0$.}

In this case, the $(2\nu)$th-order generalized KT $L_{(2\ell)i_{1}...i_{2\nu}}(q)$ appears in the independent FI (\ref{eq.pro3.1}). We can remove this $(2\nu)$th-order symmetry by introducing a sequence of an odd rank totally symmetric tensors of a higher time-dependence, that is, we add in the FI (\ref{eq.pro3.1}) the sum
\begin{equation}
\sum_{r=1,odd}^{2\nu+1} t^{2\ell+1} L_{(2\ell+1)i_{1}...i_{r}} \dot{q}^{i_{1}} ... \dot{q}^{i_{r}}+ s_{1} \frac{t^{2\ell+2}}{2\ell+2} \label{eq.prf2.1}
\end{equation}
such that $L_{(2\ell+1)i_{1}...i_{2\nu+1}}$ is a $(2\nu+1)$th-order generalized KT given by the relation
\begin{equation}
L_{(2\ell+1)i_{1}...i_{2\nu+1}}= -\frac{1}{2\ell+1} L_{(2\ell)(i_{1}...i_{2\nu}|i_{2\nu+1})}
 \label{eq.prf2.2}
\end{equation}
and the constant $s_{1}$ is defined by the relation
\begin{equation}
L_{(2\ell+1)c}Q^{c} = s_{1}. \label{eq.prf2.3}
\end{equation}

Then, from conditions (\ref{eq.pro3.3}) and (\ref{eq.pro3.5}), the restriction $N<\ell$  must be removed because now quantities of the form $L_{(2N+1)i_{1}...i_{r}}$ do exist for $N=\ell$; and we must add the condition
\begin{equation}
L_{(2\ell+1)(i_{1}...i_{r-1}|i_{r})} = (r+1) L_{(2\ell+1)i_{1}...i_{r}i_{r+1}}Q^{i_{r+1}}, \enskip r=2,4,6,...,2\nu. \label{eq.prf2.4}
\end{equation}

Observe that the condition (\ref{eq.prf2.2}) removes the original $(2\nu)$th-order symmetry defined by allowing the $(2\nu)$th-order generalized KT $L_{(2\ell)i_{1}...i_{2\nu}}(q)$ to be a totally symmetric tensor of the same rank, which is not necessarily a KT. Then, the case of the generalized KT is derived as a subcase. This shows that symmetries of an order lesser than the order of the FI can always be absorbed into a higher order term.
\bigskip

We collect the above results in the following propositions.

\begin{proposition}
\label{pro.thm1.4} For an even order $m=2\nu$, where $\nu>0$, the \textbf{complete forms} (i.e. expressions without geometric symmetries of order less than $m$) of the independent FIs (\ref{eq.pro3.1}) and (\ref{eq.pro4.1}) are the following:
\bigskip

\textbf{Integral 1.1.}
\begin{eqnarray}
J^{(2\nu,1)}_{\ell} &=& \sum_{r=1,\text{odd}}^{2\nu} \sum_{N=1}^{\ell>0} t^{2N-1} L_{(2N-1)i_{1}...i_{r}} \dot{q}^{i_{1}} ... \dot{q}^{i_{r}} +\sum_{r=1,\text{even}}^{2\nu} \sum_{N=0}^{\ell} t^{2N} L_{(2N)i_{1}...i_{r}} \dot{q}^{i_{1}} ... \dot{q}^{i_{r}} + \notag \\
&& +\sum_{N=1}^{\ell>0}L_{(2N-1)c}Q^{c} \frac{t^{2N}}{2N} +G(q) \label{eq.pro5.1}
\end{eqnarray}
where $\ell \geq0$, $L_{(2N)i_{1}...i_{2\nu}}(q)$ with $N=0,1,...,\ell$ are $\mathbf{(2\nu)}$\textbf{th-order generalized KTs} given by the relation
\begin{equation}
L_{(2N)i_{1}...i_{2\nu}}= -\frac{1}{2N} L_{(2N-1)(i_{1}...i_{2\nu-1}|i_{2\nu})}, \enskip N=1,2,...,\ell, \enskip \ell>0 \label{eq.pro5.2}
\end{equation}
and the involved quantities satisfy the conditions:
\begin{eqnarray}
G_{,i_{1}}&=& 2L_{(0)i_{1}i_{2}}Q^{i_{2}} -L_{(1)i_{1}}(\ell>0) \label{eq.pro5.3} \\
\left(L_{(2N-1)c}Q^{c} \right)_{,i_{1}}&=& 4NL_{(2N)i_{1}i_{2}} Q^{i_{2}} -2N(2N+1)L_{(2N+1)i_{1}}(N<\ell), \enskip N=1,2,...,\ell, \enskip \ell>0 \label{eq.pro5.4} \\
L_{(2N-1)(i_{1}...i_{r-1}|i_{r})} &=& (r+1) L_{(2N-1)i_{1}...i_{r}i_{r+1}}Q^{i_{r+1}} -2NL_{(2N)i_{1}...i_{r}}, \notag \\
&& N=1,2,...,\ell, \enskip r=2,4,6,...,2\nu-2, \enskip \ell>0, \enskip \nu>1 \label{eq.pro5.5} \\
L_{(2N)(i_{1}...i_{r-1}|i_{r})} &=& (r+1) L_{(2N)i_{1}...i_{r}i_{r+1}}Q^{i_{r+1}} -(2N+1)L_{(2N+1)i_{1}...i_{r}}(N<\ell), \notag \\
&& N=0,1,...,\ell, \enskip r=3,5,7,..., 2\nu-1, \enskip \nu>1. \label{eq.pro5.6}
\end{eqnarray}

\textbf{Integral 1.2.}
\begin{eqnarray}
J^{(2\nu,2)}_{\ell} &=& \sum_{r=1,\text{odd}}^{2\nu} \sum_{N=0}^{\ell} t^{2N} L_{(2N)i_{1}...i_{r}} \dot{q}^{i_{1}} ... \dot{q}^{i_{r}} +\sum_{r=1,\text{even}}^{2\nu} \sum_{N=1}^{\ell+1} t^{2N-1} L_{(2N-1)i_{1}...i_{r}} \dot{q}^{i_{1}} ... \dot{q}^{i_{r}} + \notag \\
&& +s_{0}\frac{t^{2\ell+1}}{2\ell+1} +\sum_{N=0}^{\ell-1\geq0} L_{(2N)c}Q^{c} \frac{t^{2N+1}}{2N+1} \label{eq.pro5.7}
\end{eqnarray}
where $\ell \geq0$, $L_{(2N-1)i_{1}...i_{2\nu}}(q)$ with $N=1,2,...,\ell+1$ are $\mathbf{(2\nu)}$\textbf{th-order generalized KTs} given by the relation
\begin{equation}
L_{(2N-1)i_{1}...i_{2\nu}}= -\frac{1}{2N-1} L_{(2N-2)(i_{1}...i_{2\nu-1}|i_{2\nu})}, \enskip N=1,2,...,\ell+1 \label{eq.pro5.8}
\end{equation}
and the involved quantities satisfy the conditions:
\begin{eqnarray}
\left( L_{(2N-2)c}Q^{c} \right)_{,i_{1}}&=& 2(2N-1) L_{(2N-1)i_{1}i_{2}} Q^{i_{2}} -2N(2N-1)L_{(2N)i_{1}}(N<\ell+1), \notag \\
&& N=1,2,...,\ell+1 \label{eq.pro5.9} \\
L_{(2N)(i_{1}...i_{r-1}|i_{r})} &=& (r+1) L_{(2N)i_{1}...i_{r}i_{r+1}}Q^{i_{r+1}} -(2N+1)L_{(2N+1)i_{1}...i_{r}}, \notag \\
&& N=0,1,...,\ell, \enskip r=2,4,6,...,2\nu-2, \enskip \nu>1 \label{eq.pro5.10} \\
L_{(2N-1)(i_{1}...i_{r-1}|i_{r})} &=& (r+1) L_{(2N-1)i_{1}...i_{r}i_{r+1}}Q^{i_{r+1}} -2NL_{(2N)i_{1}...i_{r}}(N<\ell+1), \notag \\
&& N=1,2,...,\ell+1, \enskip r=3,5,7,...,2\nu-1, \enskip \nu>1. \label{eq.pro5.11}
\end{eqnarray}
\end{proposition}

\begin{proposition}
\label{pro.thm1.5} For an odd order $m=2\nu+1$, where $\nu>0$, the \textbf{complete forms} (i.e. expressions without geometric symmetries of order less than $m$) of the independent FIs (\ref{eq.pro3.1}) and (\ref{eq.pro4.1}) are the following:
\bigskip

\textbf{Integral 1.1.}
\begin{eqnarray}
J^{(2\nu+1,1)}_{\ell} &=& \sum_{r=1,\text{odd}}^{2\nu+1} \sum_{N=1}^{\ell+1} t^{2N-1} L_{(2N-1)i_{1}...i_{r}} \dot{q}^{i_{1}} ... \dot{q}^{i_{r}} +\sum_{r=1,\text{even}}^{2\nu+1} \sum_{N=0}^{\ell} t^{2N} L_{(2N)i_{1}...i_{r}} \dot{q}^{i_{1}} ... \dot{q}^{i_{r}} + \notag \\
&& +s_{1}\frac{t^{2\ell+2}}{2\ell+2} +\sum_{N=1}^{\ell>0}L_{(2N-1)c}Q^{c} \frac{t^{2N}}{2N} +G(q) \label{eq.pro6.1}
\end{eqnarray}
where $\ell \geq0$, $L_{(2N-1)i_{1}...i_{2\nu+1}}(q)$ with $N=1,...,\ell+1$ are $\mathbf{(2\nu+1)}$\textbf{th-order generalized KTs} given by the relation
\begin{equation}
L_{(2N-1)i_{1}...i_{2\nu+1}}= -\frac{1}{2N-1} L_{(2N-2)(i_{1}...i_{2\nu}|i_{2\nu+1})}, \enskip N=1,2,...,\ell+1 \label{eq.pro6.2}
\end{equation}
while $s_{1}$ is a constant defined by the relation
\begin{equation}
L_{(2\ell+1)a}Q^{a} = s_{1} \label{eq.pro6.3}
\end{equation}
and the involved quantities satisfy the conditions:
\begin{eqnarray}
G_{,i_{1}}&=& 2L_{(0)i_{1}i_{2}}Q^{i_{2}} -L_{(1)i_{1}}(\ell>0) \label{eq.pro6.4} \\
\left(L_{(2N-1)c}Q^{c} \right)_{,i_{1}}&=& 4NL_{(2N)i_{1}i_{2}} Q^{i_{2}} -2N(2N+1)L_{(2N+1)i_{1}}, \enskip N=1,2,...,\ell, \enskip \ell>0 \label{eq.pro6.5} \\
L_{(2N-1)(i_{1}...i_{r-1}|i_{r})} &=& (r+1) L_{(2N-1)i_{1}...i_{r}i_{r+1}}Q^{i_{r+1}} -2NL_{(2N)i_{1}...i_{r}}(N<\ell+1), \notag \\
&& N=1,2,...,\ell+1, \enskip r=2,4,6,...,2\nu \label{eq.pro6.6} \\
L_{(2N)(i_{1}...i_{r-1}|i_{r})} &=& (r+1) L_{(2N)i_{1}...i_{r}i_{r+1}}Q^{i_{r+1}} -(2N+1)L_{(2N+1)i_{1}...i_{r}}, \notag \\
&& N=0,1,...,\ell, \enskip r=3,5,7,..., 2\nu-1, \enskip \nu>1. \label{eq.pro6.7}
\end{eqnarray}

\textbf{Integral 1.2.}
\begin{eqnarray}
J^{(2\nu+1,2)}_{\ell} &=& \sum_{r=1,\text{odd}}^{2\nu+1} \sum_{N=0}^{\ell} t^{2N} L_{(2N)i_{1}...i_{r}} \dot{q}^{i_{1}} ... \dot{q}^{i_{r}} +\sum_{r=1,\text{even}}^{2\nu+1} \sum_{N=1}^{\ell>0} t^{2N-1} L_{(2N-1)i_{1}...i_{r}} \dot{q}^{i_{1}} ... \dot{q}^{i_{r}} + \notag \\
&& +s_{0}\frac{t^{2\ell+1}}{2\ell+1} +\sum_{N=0}^{\ell-1\geq0} L_{(2N)c}Q^{c} \frac{t^{2N+1}}{2N+1} \label{eq.pro6.8}
\end{eqnarray}
where $\ell \geq0$, $L_{(2N)i_{1}...i_{2\nu+1}}(q)$ with $N=0,1,...,\ell$ are $\mathbf{(2\nu+1)}$\textbf{th-order generalized KTs} given by the relation
\begin{equation}
L_{(2N)i_{1}...i_{2\nu+1}}= -\frac{1}{2N} L_{(2N-1)(i_{1}...i_{2\nu}|i_{2\nu+1})}, \enskip N=1,2,...,\ell, \enskip \ell>0 \label{eq.pro6.9}
\end{equation}
and the involved quantities satisfy the conditions:
\begin{eqnarray}
L_{(2\ell)a}Q^{a}&=&s_{0} \label{eq.pro6.10} \\
\left( L_{(2N-2)c}Q^{c} \right)_{,i_{1}} &=& 2(2N-1) L_{(2N-1)i_{1}i_{2}} Q^{i_{2}} -2N(2N-1)L_{(2N)i_{1}}, \enskip N=1,2,...,\ell, \enskip \ell>0 \label{eq.pro6.11} \\
L_{(2N)(i_{1}...i_{r-1}|i_{r})} &=& (r+1) L_{(2N)i_{1}...i_{r}i_{r+1}}Q^{i_{r+1}} -(2N+1)L_{(2N+1)i_{1}...i_{r}}(N<\ell), \notag \\
&& N=0,1,...,\ell, \enskip r=2,4,6,...,2\nu \label{eq.pro6.12} \\
L_{(2N-1)(i_{1}...i_{r-1}|i_{r})} &=& (r+1) L_{(2N-1)i_{1}...i_{r}i_{r+1}}Q^{i_{r+1}} -2NL_{(2N)i_{1}...i_{r}}, \notag \\
&& N=1,2,...,\ell, \enskip r=3,5,7,...,2\nu-1, \enskip \ell>0, \enskip \nu>1. \label{eq.pro6.13}
\end{eqnarray}
\end{proposition}

Therefore, concerning the higher order FIs of a general autonomous dynamical system, we have the following general result.

\begin{proposition}
\label{pro.thm1.6} The autonomous (non-Riemannian in general) dynamical system (\ref{eq.tk1}) admits three independent $m$th-order FIs --autonomous or time-dependent-- with $m>0$, which are given by the following formulae: \newline
i. For $m=1$, we have the LFIs (\ref{eq.pro1.5}), (\ref{eq.pro1.10}) and (\ref{eq.pro1.7}). \newline
ii. For $m=2\nu$ with $\nu>0$,  we have the FIs (\ref{eq.pro2.7}), (\ref{eq.pro5.1}) and (\ref{eq.pro5.7}). \newline
iii. For $m=2\nu+1$ with $\nu>0$,  we have the FIs (\ref{eq.pro2.7}), (\ref{eq.pro6.1}) and (\ref{eq.pro6.8}).
\end{proposition}

In the following sections, we give applications of the above general results.

\section{Application 1: A family of 2d non-Riemannian autonomous dynamical systems}

\label{sec.nonR}

We consider two-dimensional (2d) autonomous dynamical systems of the form:
\begin{eqnarray}
\ddot{x} &=& -Q^{1}(x,y) -\Gamma^{1}_{11}(x,y)\dot{x}^{2} \label{eq.nonR1.1} \\
\ddot{y} &=& -Q^{2}(x,y) -\Gamma^{2}_{22}(x,y)\dot{y}^{2} \label{eq.nonR1.2}
\end{eqnarray}
where $q^{a}=(x,y)$ are the generalized coordinates, $Q^{1}, Q^{2}$ are the components of the generalized forces, and $\Gamma^{1}_{11}, \Gamma^{2}_{22}$ are the non-zero symmetric connection coefficients of the system (all the other connection coefficients vanish).

We will find conditions for which the dynamical system (\ref{eq.nonR1.1}) - (\ref{eq.nonR1.2}) is non-Riemannian. In that case, a kinetic metric (i.e. a regular Lagrangian or a kinetic energy) cannot be defined, and standard methods (e.g. Noether's theorem) for the determination of FIs cannot be applied. Therefore, Theorem \ref{thm1} is the only systematic method we have in order to determine the FIs of the system (\ref{eq.nonR1.1}) - (\ref{eq.nonR1.2}).

\subsection{Conditions for a non-Riemannian connection}

It is well-known that the symmetric connection coefficients $\Gamma^{a}_{bc}$ define a Riemannian connection iff there exists a (kinetic) metric $\gamma_{ab}$ with zero metricity, that is, $\gamma_{ab|c}=0$. Then,
\[
\Gamma^{a}_{bc}= \frac{1}{2}\gamma^{ad} \left( \gamma_{bd,c} +\gamma_{cd,b} -\gamma_{bc,d} \right).
\]

Question: \emph{In which cases the dynamical system (\ref{eq.nonR1.1}) - (\ref{eq.nonR1.2}) is defined over a Riemannian configuration space or, equivalently, when the associated symmetric connection is Riemannian?}

In order to answer this question, we assume that there exists a kinetic metric $\gamma_{ab}(x,y)$ such that
\begin{equation}
\gamma_{ab|c}=0 \implies \gamma_{ab,c} -\gamma_{db}\Gamma^{d}_{ac} -\gamma_{ad}\Gamma^{d}_{bc}= 0. \label{eq.nonR2}
\end{equation}
Setting $\Gamma^{1}_{11}\equiv f_{1}(x,y)$ and $\Gamma^{2}_{22}\equiv f_{2}(x,y)$, condition (\ref{eq.nonR2}) gives the following system of PDEs:
\[
\begin{cases}
\gamma_{11,x}= 2\gamma_{11}f_{1}, \enskip \gamma_{12,x}= \gamma_{12}f_{1}, \enskip \gamma_{22,x}=0 \implies \gamma_{22}= \gamma_{22}(y) \\
\gamma_{22,y}= 2\gamma_{22}f_{2}, \enskip \gamma_{12,y}= \gamma_{12}f_{2}, \enskip \gamma_{11,y}=0 \implies \gamma_{11}= \gamma_{11}(x)
\end{cases} \implies
\]
\begin{eqnarray}
\frac{d\gamma_{11}}{dx}&=& 2\gamma_{11}(x)f_{1} \label{eq.nonR3.1} \\
\frac{d\gamma_{22}}{dy}&=& 2\gamma_{22}(y)f_{2} \label{eq.nonR3.2} \\
\gamma_{12,x}&=& \gamma_{12}(x,y)f_{1} \label{eq.nonR3.3} \\
\gamma_{12,y}&=& \gamma_{12}(x,y)f_{2}. \label{eq.nonR3.4}
\end{eqnarray}

In order to have a well-defined metric $\gamma_{ab}$ (i.e. the inverse $\gamma^{ab}$ exists), it must also hold that
\begin{equation}
\det\left[\gamma_{ab}\right]\neq0 \implies \gamma_{11}\gamma_{22} -\gamma_{12}^{2} \neq0. \label{eq.nonR4}
\end{equation}

We consider the following cases:

1) Case $\gamma_{12}\neq0$.

1.1. Subcase $\gamma_{11}=\gamma_{22}=0$.

Conditions (\ref{eq.nonR3.1}), (\ref{eq.nonR3.2}) and (\ref{eq.nonR4}) are satisfied identically.

The remaining conditions (\ref{eq.nonR3.3}) and (\ref{eq.nonR3.4}) give the Riemannian connection coefficients:
\[
f_{1}= \frac{F_{,x}}{F}, \enskip f_{2}= \frac{F_{,y}}{F}
\]
where $\gamma_{12}\equiv F(x,y)$.

The associated kinetic metric is $\gamma_{ab}= F(x,y)
\left(
  \begin{array}{cc}
    0 & 1 \\
    1 & 0 \\
  \end{array}
\right)$. The constrained geodesics of this metric has been discussed for various cases of the function $F(x,y)$ in \cite{Dimakis 2019} and, more recently, in \cite{Mits Constr2022}.

1.2. Subcase $\gamma_{11}\equiv f(x)\neq0$ and $\gamma_{22}=0$.

Conditions (\ref{eq.nonR3.2}) and (\ref{eq.nonR4}) are satisfied identically.

Condition (\ref{eq.nonR3.1}) implies that $f_{1}= \frac{f_{,x}}{2f}$. Replacing $f_{1}$ in (\ref{eq.nonR3.3}), we find that $\gamma_{12}= h(y)\sqrt{f(x)}$ which when replaced into the remaining condition (\ref{eq.nonR3.4}) gives $f_{2}= \frac{h_{,y}}{h}$.

The associated kinetic metric is $\gamma_{ab}=
\left(
  \begin{array}{cc}
    f(x) & h(y)\sqrt{f} \\
    h(y)\sqrt{f} & 0 \\
  \end{array}
\right)$.

1.3. Subcase $\gamma_{22}\equiv h(y)\neq0$ and $\gamma_{11}=0$.

Conditions (\ref{eq.nonR3.1}) and (\ref{eq.nonR4}) are satisfied identically.

Condition (\ref{eq.nonR3.2}) implies that $f_{2}= \frac{h_{,y}}{2h}$. Replacing $f_{2}$ in (\ref{eq.nonR3.4}), we find that $\gamma_{12}= f(x)\sqrt{h(y)}$ which when replaced into the remaining condition (\ref{eq.nonR3.3}) gives $f_{1}= \frac{f_{,x}}{f}$.

The associated kinetic metric is $\gamma_{ab}=
\left(
  \begin{array}{cc}
    0 & f(x)\sqrt{h} \\
    f(x)\sqrt{h} & h(y) \\
  \end{array}
\right)$.

1.4. Subcase $\gamma_{11}\equiv f(x), \gamma_{22}\equiv h(y)$, and $f(x)h(y)\neq0$.

Conditions (\ref{eq.nonR3.1}) and (\ref{eq.nonR3.2}) imply that
$f_{1}=\frac{f_{,x}}{2f}$ and $f_{2}= \frac{h_{,y}}{2h}$, respectively. Replacing $f_{1}, f_{2}$ in the remaining conditions (\ref{eq.nonR3.3}), (\ref{eq.nonR3.4}), we find that $\gamma_{12}= c_{0}\sqrt{f(x)h(y)}$ where $c_{0}$ is an arbitrary non-zero constant.

The associated kinetic metric is $\gamma_{ab}=
\left(
  \begin{array}{cc}
    f(x) & c_{0}\sqrt{fh} \\
    c_{0}\sqrt{fh} & h(y) \\
  \end{array}
\right)$.

From the condition (\ref{eq.nonR4}), we find that $c_{0} \neq \pm 1$.

2) Case $\gamma_{12}=0$.

Condition (\ref{eq.nonR4}) implies that $\gamma_{11}\gamma_{22}\neq0$.

We set $\gamma_{11}\equiv f(x)$ and $\gamma_{22}\equiv h(y)$.

Conditions (\ref{eq.nonR3.3}) and (\ref{eq.nonR3.4}) are satisfied identically.

The remaining conditions (\ref{eq.nonR3.1}) and (\ref{eq.nonR3.2}) give $f_{1}=\frac{f_{,x}}{2f}$ and $f_{2}= \frac{h_{,y}}{2h}$, respectively.

The associated kinetic metric is $\gamma_{ab}=
\left(
  \begin{array}{cc}
    f(x) & 0 \\
    0 & h(y) \\
  \end{array}
\right)$.
\bigskip

We collect the above results in the following Proposition.

\begin{proposition}
\label{pro.nonR1} Autonomous dynamical systems of the form (\ref{eq.nonR1.1}) - (\ref{eq.nonR1.2}) admit a Riemannian geometry (i.e. the quantities $\Gamma^{a}_{bc}$ are Riemannian connection coefficients defined by a kinetic metric $\gamma_{ab}$) in the following cases: \newline
1) For $\Gamma^{1}_{11}= \frac{F_{,x}}{F}$ and $\Gamma^{2}_{22}= \frac{F_{,y}}{F}$, where $F(x,y)$ is a non-zero arbitrary smooth function. The associated kinetic metric is $\gamma_{ab}= F(x,y)
\left(
  \begin{array}{cc}
    0 & 1 \\
    1 & 0 \\
  \end{array}
\right)$. \newline
2) For $\Gamma^{1}_{11}= \frac{f_{,x}}{2f}$ and $\Gamma^{2}_{22}= \frac{h_{y}}{h}$, where $f(x)$ and $h(y)$ are non-zero arbitrary smooth functions. Then, $\gamma_{ab}=
\left(
  \begin{array}{cc}
    f(x) & h(y)\sqrt{f} \\
    h(y)\sqrt{f} & 0 \\
  \end{array}
\right)$. \newline
3) For $\Gamma^{1}_{11}= \frac{f_{x}}{f}$ and $\Gamma^{2}_{22}= \frac{h_{,y}}{2h}$. Then, $\gamma_{ab}=
\left(
  \begin{array}{cc}
    0 & f(x)\sqrt{h} \\
    f(x)\sqrt{h} & h(y) \\
  \end{array}
\right)$. \newline
4) For $\Gamma^{1}_{11}=\frac{f_{,x}}{2f}$ and $\Gamma^{2}_{22}= \frac{h_{,y}}{2h}$. Then, $\gamma_{ab}=
\left(
  \begin{array}{cc}
    f(x) & c_{0}\sqrt{fh} \\
    c_{0}\sqrt{fh} & h(y) \\
  \end{array}
\right)$ where the (possibly zero) constant $c_{0} \neq \pm 1$.
\end{proposition}

From Proposition \ref{pro.nonR1}, we find the following proposition.

\begin{proposition}
\label{pro.nonR2} The dynamical system (\ref{eq.nonR1.1}) - (\ref{eq.nonR1.2}) is non-Riemannian if the non-zero symmetric connection coefficients $\Gamma^{1}_{11}(x,y)$ and $\Gamma^{2}_{22}(x,y)$ satisfy the condition
\begin{equation}
\Gamma^{1}_{11,y} \neq \Gamma^{2}_{22,x}. \label{eq.nonR5}
\end{equation}
\end{proposition}

Proof:\newline
- Case 1) of Proposition \ref{pro.nonR1} implies that $\Gamma^{1}_{11}= (\ln F)_{,x}$ and $\Gamma^{2}_{22}= (\ln F)_{,y}$. Taking the integrability condition $(\ln F)_{,xy}= (\ln F)_{,yx}$, we find $\Gamma^{1}_{11,y}= \Gamma^{2}_{22,x}$. Therefore, for a non-Riemannian connection, the condition (\ref{eq.nonR5}) is required. \newline
- For the remaining cases 2) - 4) of Proposition \ref{pro.nonR1}, we observe that $\Gamma^{1}_{11}(x)$ and $\Gamma^{2}_{22}(y)$. This implies that $\Gamma^{1}_{11,y}= \Gamma^{2}_{22,x}= 0$, which is a subcase of the condition $\Gamma^{1}_{11,y}= \Gamma^{2}_{22,x}$. Therefore, for a non-Riemannian connection, we find again the condition (\ref{eq.nonR5}).

\subsection{LFIs for the non-Riemannian dynamical system (\ref{eq.nonR1.1}) - (\ref{eq.nonR1.2})}

We assume that the connection of the autonomous dynamical system (\ref{eq.nonR1.1}) - (\ref{eq.nonR1.2}) is non-Riemannian; therefore, the condition (\ref{eq.nonR5}) applies.

Applying Theorem \ref{thm1} for $m=1$ (LFIs) and $n=0$ (zero degree of time-dependence), we find for the considered dynamical system the autonomous LFI
\begin{equation}
I= L_{1}(x,y)\dot{x} +L_{2}(x,y)\dot{y} +s_{0}t \label{eq.nonR6.1}
\end{equation}
where $L_{a}(x,y)$ is a generalized KV of the connection $\Gamma^{a}_{bc}(x,y)$ defined by the dynamical equations (\ref{eq.nonR1.1}) - (\ref{eq.nonR1.2}) and $s_{0}$ is a constant defined by the condition
\begin{equation}
L_{1}Q^{1} +L_{2}Q^{2}= s_{0}. \label{eq.nonR6.2}
\end{equation}

The generalized KV condition $L_{(a|b)}=0$ implies the following system of PDEs:
\begin{eqnarray}
L_{1,x} -L_{1}\Gamma^{1}_{11} &=& 0 \label{eq.nonR6.3} \\
L_{2,y} -L_{2}\Gamma^{2}_{22} &=& 0 \label{eq.nonR6.4} \\
L_{1,y} +L_{2,x} &=& 0. \label{eq.nonR6.5}
\end{eqnarray}

We have an overdetermined system of four PDEs (\ref{eq.nonR6.2}) - (\ref{eq.nonR6.5}), which has six unknown functions $L_{a}(x,y)$, $Q^{a}(x,y)$, $\Gamma^{1}_{11}(x,y)$, $\Gamma^{2}_{22}(x,y)$ and one free parameter $s_{0}$. Therefore, in order to solve it, we should fix either the dynamics of the system (i.e. the generalized forces $Q^{a}$) or the non-Riemannian geometry of the system (i.e. the non-Riemannian connection coefficients; it holds that $\Gamma^{1}_{11,y} \neq \Gamma^{2}_{22,x}$).

\subsubsection{Two linearly coupled harmonic oscillators with a non-Riemannian quadratic damping term}

As a first application, we fix the generalized forces $Q^{a}$. We assume that
\begin{equation}
Q^{a}=
\left(
  \begin{array}{c}
    kx -py \\
    ky +px \\
  \end{array}
\right) \label{eq.nonR6.6}
\end{equation}
where $k, p$ are arbitrary non-zero constants.

For the choice (\ref{eq.nonR6.6}), the dynamical system (\ref{eq.nonR1.1}) - (\ref{eq.nonR1.2}) becomes:
\begin{eqnarray}
\ddot{x} &=& -kx +py -\Gamma^{1}_{11}(x,y)\dot{x}^{2} \label{eq.nonR7.1} \\
\ddot{y} &=& -ky -px -\Gamma^{2}_{22}(x,y)\dot{y}^{2} \label{eq.nonR7.2}
\end{eqnarray}
which describes two linearly coupled harmonic oscillators with a non-Riemannian (i.e. $\Gamma^{1}_{11,y} \neq \Gamma^{2}_{22,x}$) quadratic damping term.

Replacing $Q^{a}$ from (\ref{eq.nonR6.6}), equation (\ref{eq.nonR6.2}) gives
\begin{equation}
L_{1}= -\frac{ky +px}{kx-py}L_{2} +\frac{s_{0}}{kx-py}= \frac{1}{py -kx} \left[ (ky+px)L_{2} -s_{0} \right] \label{eq.nonR8}
\end{equation}
which when substituted into (\ref{eq.nonR6.5}) implies that
\begin{equation}
L_{2}= (py -kx) F_{1}\left( p(y^{2}-x^{2}) -2kxy\right) -\frac{s_{0}x}{p(y^{2}-x^{2}) -2kxy} \label{eq.nonR9}
\end{equation}
where $F_{1}\left( p(y^{2}-x^{2}) -2kxy\right)$ is an arbitrary smooth function of its argument.

Replacing (\ref{eq.nonR9}) in (\ref{eq.nonR8}), we find
\begin{equation}
L_{1}= (ky+px) F_{1}\left( p(y^{2}-x^{2}) -2kxy\right) +\frac{s_{0}x(ky+px)}{(kx -py)\left[ p(y^{2}-x^{2}) -2kxy\right]} +\frac{s_{0}}{kx-py}. \label{eq.nonR10}
\end{equation}

Using (\ref{eq.nonR9}) and (\ref{eq.nonR10}), the remaining PDEs (\ref{eq.nonR6.3}) and (\ref{eq.nonR6.4}) determine the connection coefficients as follows:
\begin{equation}
\Gamma^{1}_{11}= \frac{L_{1,x}}{L_{1}}, \enskip \Gamma^{2}_{22}= \frac{L_{2,y}}{L_{2}}. \label{eq.nonR11}
\end{equation}
In order to have a non-Riemannian connection, the condition (\ref{eq.nonR5}) must be satisfied. Therefore,
\begin{equation}
\Gamma^{1}_{11,y} \neq \Gamma^{2}_{22,x} \implies \left( \ln |L_{1}| \right)_{,xy} \neq \left( \ln |L_{2}| \right)_{,yx} \implies \left( \ln \left| \frac{L_{1}}{L_{2}} \right| \right)_{,xy} \neq 0 \label{eq.nonR11.0}
\end{equation}
which leads to restrictions among the parameters $k, p, s_{0}$ and the function $F_{1}$.

We note that the vector $L_{a}$ whose components are given by the relations (\ref{eq.nonR9}) and (\ref{eq.nonR10}) is a generalized KV of the connection (\ref{eq.nonR11}).

Finally, replacing (\ref{eq.nonR11}) in the dynamical equations (\ref{eq.nonR7.1}) - (\ref{eq.nonR7.2}), we find the family of dynamical systems:
\begin{eqnarray}
\ddot{x} &=& -kx +py -\frac{L_{1,x}}{L_{1}}\dot{x}^{2} \label{eq.nonR11.1} \\
\ddot{y} &=& -ky -px -\frac{L_{2,y}}{L_{2}}\dot{y}^{2} \label{eq.nonR11.2}
\end{eqnarray}
parameterized by the constant $s_{0}$ and the function $F_{1}$, which admits the time-dependent LFI (\ref{eq.nonR6.1})

In order to get a specific example, we fix the parameters $s_{0}$ and $F_{1}$ as follows:

- Case $s_{0}=0$ and $F_{1}=1$.

Then, equations (\ref{eq.nonR9}) and (\ref{eq.nonR10}) give the generalized KV
\begin{equation}
L_{a}=
\left(
  \begin{array}{c}
    ky+px \\
    py-kx \\
  \end{array}
\right) \label{eq.nonR12}
\end{equation}
which when replaced into the relations (\ref{eq.nonR11}) determines the connection coefficients:
\begin{equation}
\Gamma^{1}_{11}= \frac{p}{ky+px}, \enskip \Gamma^{2}_{22}= \frac{p}{py-kx}. \label{eq.nonR13}
\end{equation}
In order to have a non-Riemannian connection, the quantities (\ref{eq.nonR13}) must satisfy the condition (\ref{eq.nonR11.0}). We compute:
\[
\Gamma^{1}_{11,y}= -\frac{kp}{(ky+px)^{2}} \neq \Gamma^{2}_{22,x}= \frac{kp}{(py-kx)^{2}} \implies
\]
\[
(py-kx)^{2} +(ky+px)^{2} \neq0 \implies (k^{2}+p^{2})(x^{2}+y^{2}) \neq0 \implies k\neq \pm ip.
\]
Therefore, \emph{the connection (\ref{eq.nonR13}) is non-Riemannian only when $k\neq \pm ip$}.

Using (\ref{eq.nonR12}), the dynamical system (\ref{eq.nonR11.1}) - (\ref{eq.nonR11.2}) becomes:
\begin{eqnarray}
\ddot{x} &=& -kx +py -\frac{p}{ky+px}\dot{x}^{2} \label{eq.nonR14.1} \\
\ddot{y} &=& -ky -px -\frac{p}{py-kx}\dot{y}^{2} \label{eq.nonR14.2}
\end{eqnarray}
and the associated autonomous LFI (\ref{eq.nonR6.1}) is
\begin{equation}
I_{1}= (ky +px)\dot{x} +(py-kx)\dot{y}. \label{eq.nonR14.3}
\end{equation}
\bigskip

\underline{Remark:} In the case that $k=\pm ip$, the connection coefficients (\ref{eq.nonR13}) reduce to the Riemannian connection:
\[
\Gamma^{1}_{11\pm}= \frac{1}{x\pm iy}, \enskip \Gamma^{2}_{22\pm}= \frac{1}{y\mp ix}
\]
which, according to Proposition \ref{pro.nonR1}, is associated with a metric of the form 1). Indeed, for an arbitrary function $F(x,y)$, we have:
\[
\begin{cases}
(\ln F_{\pm})_{,x}= \frac{1}{x\pm iy} \\
(\ln F_{\pm})_{,y}= \frac{1}{y\mp ix}
\end{cases} \implies
F_{+}= \frac{x^{2}+y^{2}}{y+ix}, \enskip F_{-}=y+ix
\]
and the corresponding metric is $\gamma_{ab}= F_{\pm}
\left(
  \begin{array}{cc}
    0 & 1 \\
    1 & 0 \\
  \end{array}
\right)$.

\section{Application 2: The QFIs of a non-Riemannian dynamical system}

\label{sec.andro}

Consider the dynamical system\footnote{This dynamical system has been suggested to us by Dr. Andronikos Paliathanasis.}:
\begin{eqnarray}
\ddot{u} &=& -\frac{8\beta }{u^{3}}\left( u\dot{u}\dot{w} -w\dot{u}^{2}\right) -\frac{1}{u^{2}} \label{eq.aham8a} \\
\ddot{w} &=& -\frac{4\beta }{u^{3}}\left( u\dot{w}^{2} -4w\dot{u}\dot{w}\right) +\frac{2w}{u^{3}} \label{eq.aham8b}
\end{eqnarray}
where $\beta$ is an arbitrary real constant. This system is autonomous holonomic of the form (\ref{eq.tk1}) with variables
\[
q^{a}=
\left(
  \begin{array}{c}
    u \\
    w \\
  \end{array}
\right), \enskip
Q^{a}= \frac{1}{u^{2}}
\left(
  \begin{array}{c}
    1 \\
    -\frac{2w}{u} \\
  \end{array}
\right).
\]
The symmetric connection coefficients are read from the dynamical equations and are:
\begin{equation}
\Gamma^{1}_{22}=\Gamma^{2}_{11}=0, \enskip \Gamma^{1}_{11}= \Gamma^{2}_{12}= -8\beta\frac{w}{u^{3}}, \enskip \Gamma^{1}_{12}= \Gamma^{2}_{22}= \frac{4\beta}{u^{2}}. \label{eq.conne}
\end{equation}

The non-zero components of the curvature tensor $R^{a}{}_{bcd}= \Gamma^{a}_{bd,c} -\Gamma^{a}_{bc,d} +\Gamma^{a}_{sc}\Gamma^{s}_{bd} -\Gamma^{a}_{sd}\Gamma^{s}_{bc}$ are:
\[
R^{1}{}_{112}= R^{2}{}_{221} = -R^{2}{}_{212} = -R^{1}{}_{121}= -\frac{32b^{2}w}{u^{5}}, \enskip R^{2}{}_{112}= -R^{2}{}_{121}= \frac{24bw}{u^{4}}.
\]

Solving the generalized KT condition $C_{(ab|c)}=0$, we find that the connection (\ref{eq.conne}) admits only the second order generalized KT
\begin{equation}
C_{ab}= k e^{\frac{12\beta w}{u^{2}}}
\left(
  \begin{array}{cc}
    0 & 1 \\
    1 & 0 \\
  \end{array}
\right) \label{eq.KT1}
\end{equation}
where $k$ is an arbitrary constant.

Solving the generalized Killing vector (KV) condition $L_{(a|b)}=0$, we find $L_{a}=0$; therefore, generalized KVs do not exist.

Moreover, it can be shown that non-zero vectors $B_{a}$ which generate reducible generalized KTs of the form $B_{(a|b)}$ do not exist as well.

Applying Theorem \ref{thm1} for $m=2$, we find that the system admits only one QFI which is the
\begin{equation}
I= e^{\frac{12\beta w}{u^{2}}} \left( \dot{u}\dot{w} +\frac{1}{12\beta} \right). \label{eq.qint1.5}
\end{equation}
To prove that the given system is integrable, one needs one more independent autonomous FI of higher order in involution.

\section{Application 3: A new superintegrable potential which admits time-dependent QFIs}

\label{sec.appl2}

In \cite{Evans 1990}, using the separability of the corresponding Hamilton-Jacobi equation in more than two coordinate systems, all minimally and maximally superintegrable potentials in the Euclidean space $E^{3}$ that admit autonomous QFIs are determined. We extend this result to the case where time-dependent FIs are considered.

Applying Theorem \ref{thm1} for $m=2$ (QFIs), $q^{a}=(x,y,z)$, $\Gamma^{a}_{bc}=0$ and $Q^{a}=V^{a,}$, where $V(x,y,z)$ denotes the potential, we find the new maximally superintegrable potential in $E^{3}$
\begin{equation}
V(x,y,z)= -\frac{\lambda^{2}}{2} R^{2} +\frac{kx}{y^{2}R} +\frac{c_{1}}{y^{2}} -\frac{\lambda^{2}}{8}z^{2} +\frac{c_{2}}{z^{2}} \label{eq.evans.1}
\end{equation}
where $\lambda\neq0, k, c_{1}, c_{2}$ are arbitrary constants, and $R= \sqrt{x^{2} +y^{2}}$.

The potential (\ref{eq.evans.1}) admits the following independent (autonomous and time-dependent) QFIs:
\begin{eqnarray}
I_{1}&=& \frac{1}{2} \left( \dot{x}^{2} +\dot{y}^{2} +\dot{z}^{2} \right) -\frac{\lambda^{2}}{2} R^{2} +\frac{kx}{y^{2}R} +\frac{c_{1}}{y^{2}} -\frac{\lambda^{2}}{8}z^{2} +\frac{c_{2}}{z^{2}} \label{eq.evans.2.1} \\
I_{2}&=& \frac{1}{2}M_{3}^{2} + \frac{(kR +c_{1}x)x}{y^{2}} \label{eq.evans.2.2} \\
I_{3}&=& \frac{1}{2}\dot{z}^{2} -\frac{\lambda^{2}}{8}z^{2} +\frac{c_{2}}{z^{2}} \label{eq.evans.2.3} \\
I_{4}&=& e^{\lambda t} \left[ M_{3} (\dot{y} -\lambda y) +\frac{2c_{1}x}{y^{2}} +\frac{k(y^{2} +2x^{2})}{y^{2}R} \right] \label{eq.evans.2.4} \\
I_{5}&=& e^{\lambda t} \left[ \left( \dot{z} -\frac{\lambda}{2}z \right)^{2} +\frac{2c_{2}}{z^{2}} \right]. \label{eq.evans.2.5}
\end{eqnarray}
The QFI $I_{1}$ is the Hamiltonian of the system and the vector $M_{i}= \left( y\dot{z} -z\dot{y}, z\dot{x} -x\dot{z}, x\dot{y} -y\dot{x} \right)$ with $i=1,2,3$ is the angular momentum.

\section{Application 4: A new superintegrable separable potential which admits an autonomous CFI}

\label{sec.appl3}

It is well-known that separable Newtonian potentials of the form $V(x,y)= F_{1}(x) +F_{2}(y)$ admit the QFIs $J_{1}= \frac{1}{2}\dot{x}^{2} +F_{1}(x)$ and $J_{2}= \frac{1}{2}\dot{y}^{2} +F_{2}(y)$, where $F_{1}$ and $F_{2}$ are arbitrary smooth functions of their arguments. The question is if there are functions $F_{1}$ and $F_{2}$ for which the corresponding potential $V(x,y)$ is superintegrable.

A partial answer to this problem has been given in \cite{Gravel 2004} by considering autonomous CFIs as the third FI. One of the third order superintegrable potentials found in \cite{Gravel 2004} is the
\begin{equation}
V(x,y) = c_{1}y^{2} +F(x) \label{eq.gravel.1}
\end{equation}
where $c_{1}$ is an arbitrary non-zero constant and $F(x)$ is an arbitrary smooth function satisfying the condition
\begin{equation}
k_{2}x^{2} +4k_{1}^{2} +\left( 9F -c_{1}x^{2} \right) \left( F -c_{1}x^{2} \right)^{3} -4k_{1} \left( F -c_{1}x^{2} \right) \left( 3F +c_{1}x^{2} \right)= 0 \label{eq.gravel.1.2}
\end{equation}
where $k_{1}$ and $k_{2}$ are arbitrary constants.

Using Theorem \ref{thm1}, we generalize the above result and determine a class of superintegrable potentials which contains the potential (\ref{eq.gravel.1}) as a special case.

We apply Theorem \ref{thm1} for $m=3$ (CFIs), $q^{a}=(x,y)$, $\Gamma^{a}_{bc}=0$ and $Q^{a}= V^{,a}$, where $V(x,y)$ denotes potentials of the form $V= F_{1}(x) +F_{2}(y)$. We find that the potential (\ref{eq.gravel.1}) is superintegrable due to the three independent autonomous FIs:
\begin{eqnarray}
I_{1}&=& \frac{1}{2}\dot{x}^{2} +F(x) \label{eq.gravel.3.1} \\
I_{2}&=& \frac{1}{2}\dot{y}^{2} +c_{1}y^{2} \label{eq.gravel.3.2} \\
I_{3}&=& L\dot{x}^{2} -\left( 3yF -c_{1}x^{2}y +k_{3}y \right) \dot{x} +\frac{F'}{2c_{1}}\left( 3F -c_{1}x^{2} +k_{3} \right)\dot{y} \label{eq.gravel.3.3}
\end{eqnarray}
where $L\equiv x\dot{y} -y\dot{x}$ is the angular momentum, $F'\equiv \frac{dF}{dx}$ and the function $F(x)$ satisfies the condition
\begin{eqnarray}
0&=& k_{2}x^{2} +4k_{1}^{2} +\left( 9F -c_{1}x^{2} \right) \left( F -c_{1}x^{2} \right)^{3} -4k_{1} \left( F -c_{1}x^{2} \right) \left( 3F +c_{1}x^{2} \right) + \notag \\
&& +4k_{3} \left( 3F -c_{1}x^{2} \right) \left( F -c_{1}x^{2} \right)^{2} +4k_{3}^{2} \left( F -c_{1}x^{2} \right)^{2} -\frac{8k_{1}k_{3}}{3} \left( 3F -c_{1}x^{2} \right) \label{eq.gravel.2}
\end{eqnarray}
where $k_{1}, k_{2}, k_{3}$ are arbitrary constants. We note that for $k_{3}=0$ condition (\ref{eq.gravel.2}) reduces to condition (\ref{eq.gravel.1.2}). Therefore, the superintegrable potential (C.6) of \cite{Gravel 2004} is a subcase of (\ref{eq.gravel.1}).

\section{Conclusions}

\label{sec.conclusions}

We draw the following conclusions: \newline
a) We have developed a direct systematic method to compute the $m$th-order FIs of the autonomous (in general non-Riemannian) dynamical systems (\ref{eq.tk1}) in terms of the `symmetries' of the geometric objects (symmetric connection or kinetic metric, depending on the case) defined by the dynamical equations. \newline
b) This method applies to non-Riemanian geometries with a symmetric connection. It has been shown that the $m$th-order FIs require the generalized KTs and KVs defined by the symmetric connection $\Gamma _{bc}^{a}(q)$. The case of a Riemannian connection is a special case, where the $m$th-order FIs can be related to a gauged weak generalized Noether symmetry by means of the Inverse Noether Theorem.\newline
c) The system of PDEs (\ref{eq.veldep4.1}) - (\ref{eq.veldep4.4}) resulting from the condition $\frac{dI^{(m)}}{dt}=0$ along the dynamical equations consists of two parts: A geometric part (eqs. (\ref{eq.veldep4.1}) and (\ref{eq.veldep4.2}) ) common to all systems which share the same connection; and a dynamical part (eqs. (\ref{eq.veldep4.3}) and (\ref{eq.veldep4.4}) ) which includes the generalized forces $Q^{a}$ of the specific system. \newline
d) We determined the condition which the connection coefficients must satisfy in order the 2d dynamical systems (\ref{eq.nonR1.1}) - (\ref{eq.nonR1.2}) to be non-Riemannian.

Obviously, Theorem \ref{thm1} provides a new systematic way to determine the higher order FIs, autonomous and time-dependent, of autonomous (in general non-Riemannian) dynamical systems of the form (\ref{eq.tk1}).

\section*{Appendix}

We assume that the totally symmetric tensor quantities
\begin{equation}
M_{i_{1}...i_{r}}(t,q)=\sum_{N_{r}=0}^{n_{r}} L_{(N_{r})i_{1}...i_{r}}(q)t^{N_{r}},\enskip r=1,2,...,m, \enskip m\geq1 \label{eq.aspm}
\end{equation}
where $L_{(N_{r})i_{1}...i_{r}}(q)$, $N_{r}=0,1,...,n_{r}$, are arbitrary $r$-rank totally symmetric tensors and $n_{r}\geq0$ is the degree of the polynomial associated with the $r$-rank tensor $M_{i_{1}...i_{r}}(t,q)$. We note that the degrees $n_{r}$ of the above polynomial expressions of $t$ may be infinite.

We consider the following cases.
\bigskip

\underline{\textbf{I. Case with $n$ finite.}}

Substituting (\ref{eq.aspm}) in the system of PDEs (\ref{eq.veldep4.1}) - (\ref{eq.veldep4.4}), we obtain the following system of polynomial equations in $t$:
\begin{eqnarray}
0&=& \sum_{N=0}^{n} L_{(N)(i_{1}...i_{m}|i_{m+1})}t^{N} \label{eq.nh1.0} \\
0&=& M_{,t} -\sum^{n}_{N=0} L_{(N)i_{1}}Q^{i_{1}} t^{N} \label{eq.nh1.1} \\
0 &=& M_{,i_{1}} +\sum^{n>0}_{N=1}\left[ NL_{(N)i_{1}} -2L_{(N-1)i_{1}i_{2}}Q^{i_{2}}\right] t^{N-1} -2L_{(n)i_{1}i_{2}}Q^{i_{2}}t^{n}, \enskip m>1 \label{eq.nh1.2} \\
0&=& \sum^{n-1\geq0}_{N=0} \left[ (N+1) L_{(N+1)i_{1}...i_{r}} +L_{(N)(i_{1}...i_{r-1}|i_{r})} -(r+1)L_{(N)i_{1}...i_{r}i_{r+1}} Q^{i_{r+1}} \right] t^{N} + \notag \\
&& +\left[ L_{(n)(i_{1}...i_{r-1}|i_{r})} -(r+1)L_{(n)i_{1}...i_{r}i_{r+1}} Q^{i_{r+1}} \right] t^{n}, \enskip r= 2,3,...,m-1, \enskip m>2 \label{eq.nh1.3} \\
0&=& M_{,i_{1}} +\sum^{n>0}_{N=1}NL_{(N)i_{1}} t^{N-1}, \enskip m=1 \label{eq.nh1.4} \\
0&=& \sum^{n-1\geq0}_{N=0} \left[ (N+1)L_{(N+1)i_{1}...i_{m}} +L_{(N)(i_{1}...i_{m-1}|i_{m})} \right]t^{N} +L_{(n)(i_{1}...i_{m-1}|i_{m})}t^{n}, \enskip m>1 \label{eq.nh1.5}
\end{eqnarray}
where --without loss of generality-- the polynomial expressions (\ref{eq.aspm}) of $t$ are assumed to be of the same degree, that is, $n=n_{r}$ for all values of $r$. All the results with $n \neq n_{r}$ are derived as subcases from the case $n=n_{r}$. We note also that: Equation (\ref{eq.nh1.0}) is derived from (\ref{eq.veldep4.1}), (\ref{eq.nh1.1}) from (\ref{eq.veldep4.4}), (\ref{eq.nh1.2}) from (\ref{eq.veldep4.3}) for $r=1$, (\ref{eq.nh1.3}) from (\ref{eq.veldep4.3}) for $r=2,3,...,m-1$, (\ref{eq.nh1.4}) from (\ref{eq.veldep4.2}) for $m=1$, and (\ref{eq.nh1.5}) from (\ref{eq.veldep4.2}) for $m>1$.

Equation (\ref{eq.nh1.0}) implies that the quantities $L_{(N)i_{1}...i_{m}}$ with $N=0,1,...,n$ are $m$th-order generalized KTs. For $m=1$, $L_{(N)i_{1}}$ are generalized KVs.

Integrating equation (\ref{eq.nh1.1}), we find that
\begin{equation}
M= \sum^{n}_{N=0} L_{(N)i_{1}}Q^{i_{1}} \frac{t^{N+1}}{N+1} +G(q) \label{eq.nh2.1}
\end{equation}
where $G(q)$ is an arbitrary smooth function. \emph{We note that the integrability conditions of the scalar $M(t,q)$ have been replaced by the integrability conditions $G_{,[ab]}=0$ of the function $G(q)$.}

Replacing $M$ from (\ref{eq.nh2.1}), equation (\ref{eq.nh1.2}) gives:
\begin{align*}
0=& G_{,i_{1}} +L_{(1)i_{1}}(n>0) -2L_{(0)i_{1}i_{2}}Q^{i_{2}} + \\
& +\sum^{n-1\geq1}_{N=1} \frac{1}{N} \left[ \left(L_{(N-1)c}Q^{c}\right)_{,i_{1}} +N(N+1)L_{(N+1)i_{1}} -2NL_{(N)i_{1}i_{2}}Q^{i_{2}} \right] t^{N} + \\
& +\frac{1}{n} \underbrace{\left[ \left(L_{(n-1)c}Q^{c}\right)_{,i_{1}} -2nL_{(n)i_{1}i_{2}}Q^{i_{2}} \right]}_{n>0} t^{n} +\left(L_{(n)c}Q^{c}\right)_{,i_{1}} \frac{t^{n+1}}{n+1}, \enskip m>1 \implies
\end{align*}
\begin{eqnarray}
G_{,i_{1}} &=& 2L_{(0)i_{1}i_{2}}Q^{i_{2}} -L_{(1)i_{1}}(n>0), \enskip m>1 \label{eq.nh3.1} \\
\left(L_{(k-1)c}Q^{c}\right)_{,i_{1}} &=& 2kL_{(k)i_{1}i_{2}}Q^{i_{2}} -k(k+1)L_{(k+1)i_{1}}, \enskip k=1,2,...,n-1, \enskip n>1, \enskip m>1 \label{eq.nh3.2} \\
\left(L_{(n-1)c}Q^{c}\right)_{,i_{1}} &=& 2nL_{(n)i_{1}i_{2}}Q^{i_{2}}, \enskip n>0, \enskip m>1 \label{eq.nh3.3} \\
L_{(n)i_{1}}Q^{i_{1}} &=& s_{0}, \enskip m>1 \label{eq.nh3.4}
\end{eqnarray}
where $s_{0}$ is an arbitrary constant. \emph{The notation $L_{(1)i_{1}}(n>0)$ indicates that the vector $L_{(1)i_{1}}$ exists only when the degree of the polynomial $n>0$, that is, when $n=0$, the vector $L_{(1)i_{1}}$ vanishes}.

Replacing (\ref{eq.nh2.1}) in (\ref{eq.nh1.4}), we find the following conditions:
\begin{eqnarray}
G_{,i_{1}} &=& -L_{(1)i_{1}}(n>0), \enskip m=1 \label{eq.nh4.1} \\
\left(L_{(k-1)c}Q^{c}\right)_{,i_{1}} &=& -k(k+1)L_{(k+1)i_{1}}, \enskip k=1,2,...,n-1, \enskip n>1, \enskip m=1 \label{eq.nh4.2} \\
L_{(n-1)i_{1}}Q^{i_{1}} &=& s_{1}, \enskip n>0, \enskip m=1 \label{eq.nh4.3} \\
L_{(n)i_{1}}Q^{i_{1}} &=& s_{0}, \enskip m=1 \label{eq.nh4.4}
\end{eqnarray}
where $s_{1}$ is an arbitrary constant.

We observe that conditions (\ref{eq.nh4.1}) - (\ref{eq.nh4.4}) are emerged from conditions (\ref{eq.nh3.1}) - (\ref{eq.nh3.4}) if we rewrite the latter in the following compact form:
\begin{eqnarray}
G_{,i_{1}} &=& 2L_{(0)i_{1}i_{2}}(m>1)Q^{i_{2}} -L_{(1)i_{1}}(n>0) \label{eq.nh5.1} \\
\left(L_{(k-1)c}Q^{c}\right)_{,i_{1}} &=& 2kL_{(k)i_{1}i_{2}}(m>1)Q^{i_{2}} -k(k+1)L_{(k+1)i_{1}}, \enskip k=1,2,...,n-1, \enskip n>1 \label{eq.nh5.2} \\
\left(L_{(n-1)c}Q^{c}\right)_{,i_{1}} &=& 2nL_{(n)i_{1}i_{2}}(m>1)Q^{i_{2}}, \enskip n>0 \label{eq.nh5.3} \\
L_{(n)i_{1}}Q^{i_{1}} &=& s_{0}. \label{eq.nh5.4}
\end{eqnarray}
\emph{The notation $L_{(0)i_{1}i_{2}}(m>1)$ indicates that the quantities $L_{(0)i_{1}i_{2}}$ exist only when the degree of the FI $m>1$, that is, when $m=1$, the quantities $L_{(0)i_{1}i_{2}}$ vanish.}

Conditions (\ref{eq.nh5.2}) and (\ref{eq.nh5.3}) are written compactly as follows:
\begin{equation}
\left(L_{(k-1)c}Q^{c}\right)_{,i_{1}} = 2kL_{(k)i_{1}i_{2}}(m>1)Q^{i_{2}} -k(k+1)L_{(k+1)i_{1}}(k<n), \enskip k=1,2,...,n, \enskip n>0. \label{eq.nh5.5}
\end{equation}
\emph{The notation $L_{(k+1)i_{1}}(k<n)$ indicates that the vector $L_{(k+1)i_{1}}$ exists only when $k<n$, that is, if $k\geq n$, the vector $L_{(k+1)i_{1}}$ vanishes.}

Equation (\ref{eq.nh1.3}) implies that:
\begin{eqnarray}
L_{(k)(i_{1}...i_{r-1}|i_{r})} &=& (r+1)L_{(k)i_{1}...i_{r}i_{r+1}} Q^{i_{r+1}} -(k+1) L_{(k+1)i_{1}...i_{r}}, \notag \\
&& k=0,1,..., n-1, \enskip r= 2,3,...,m-1, \enskip n>0, \enskip m>2 \label{eq.nh6.1} \\
L_{(n)(i_{1}...i_{r-1}|i_{r})} &=& (r+1)L_{(n)i_{1}...i_{r}i_{r+1}} Q^{i_{r+1}}, \enskip r= 2,3,...,m-1, \enskip m>2. \label{eq.nh6.2}
\end{eqnarray}

Conditions (\ref{eq.nh6.1}) and (\ref{eq.nh6.2}) are written compactly as follows:
\begin{eqnarray}
L_{(k)(i_{1}...i_{r-1}|i_{r})} &=& (r+1)L_{(k)i_{1}...i_{r}i_{r+1}} Q^{i_{r+1}} -(k+1) L_{(k+1)i_{1}...i_{r}}(k<n), \notag \\
&& k=0,1,..., n, \enskip r= 2,3,...,m-1, \enskip m>2. \label{eq.nh7}
\end{eqnarray}
\emph{The notation $L_{(k+1)i_{1}...i_{r}}(k<n)$ indicates that the quantities $L_{(k+1)i_{1}...i_{r}}$ exist only when $k<n$, that is, if $k\geq n$, the quantities $L_{(k+1)i_{1}...i_{r}}$ vanish.}

Equation (\ref{eq.nh1.5}) implies that the quantities $L_{(n)i_{1}...i_{m-1}}$ with $m>1$ are the components of an $(m-1)$th-order generalized KT and the $m$th-order generalized KTs
\begin{equation}
L_{(k)i_{1}...i_{m}} = -\frac{1}{k} L_{(k-1)(i_{1}...i_{m-1}|i_{m})}, \enskip k=1, 2, ..., n, \enskip n>0, \enskip m>1. \label{eq.nh8}
\end{equation}

The $m$th-order FI (\ref{FI.5}) is
\begin{equation}
I^{(m)}_{n}= \sum_{r=1}^{m} \left( \sum^{n}_{N=0}L_{(N)i_{1}...i_{r}} t^{N}\right) \dot{q}^{i_{1}} ... \dot{q}^{i_{r}} +s_{0}\frac{t^{n+1}}{n+1} +\sum^{n>0}_{N=1} L_{(N-1)c}Q^{c} \frac{t^{N}}{N} +G(q) \label{eq.nh9}
\end{equation}
where $m\geq1$, $n\geq0$, $L_{(N)i_{1}...i_{m}}(q)$ with $N=0, 1, ..., n$ are $m$th-order generalized KTs satisfying the condition (\ref{eq.nh8}), $L_{(n)i_{1}...i_{m-1}}(q)$ with $m>1$ is an $(m-1)$th-order generalized KT and the constant $s_{0}$ is given by (\ref{eq.nh5.4}). The function $G(q)$ and the totally symmetric tensors $L_{(N)i_{1}...i_{r}}(q)$ satisfy the conditions (\ref{eq.nh5.1}), (\ref{eq.nh5.5}) and (\ref{eq.nh7}).

The notation $I^{(m)}_{n}$ means the $m$th-order FI (upper index) with time-dependence $n$ (lower index). For example, the $I^{(2)}_{n}$ is a QFI whose coefficients are expressed as polynomials of $t$ of degree fixed by $n$.
\bigskip

\underline{\textbf{II. Case with $n$ infinite.}}

The polynomial expressions (\ref{eq.aspm}) as $n_{r}=n \to \infty$ turn into the infinite sum (series)
\begin{equation}
M_{i_{1}...i_{r}}(t,q)= \sum_{N=0}^{\infty} L_{(N)i_{1}...i_{r}}(q) t^{N}, \enskip r=1,2,...,m, \enskip m\geq1. \label{eq.nh10.1}
\end{equation}
It is found that new results --different from those found with $n$ finite-- are derived in the case that
\begin{equation}
L_{(N)i_{1}...i_{r}}(q)= \frac{\lambda_{r}^{N}}{N!} L_{i_{1}...i_{r}}(q) \label{eq.nh10.2}
\end{equation}
where $\lambda_{r}$ are arbitrary non-zero constants and $L_{i_{1}...i_{r}}(q)$ are $r$-rank totally symmetric tensors.

Replacing (\ref{eq.nh10.2}) in (\ref{eq.nh10.1}), we find
\begin{equation}
M_{i_{1}...i_{r}}(t,q)= L_{i_{1}...i_{r}}(q) \sum_{N=0}^{\infty} \frac{(\lambda_{r}t)^{N}}{N!} = e^{\lambda_{r}t} L_{i_{1}...i_{r}}(q). \label{eq.nh10.3}
\end{equation}

Substituting (\ref{eq.nh10.3}) in the system of PDEs (\ref{eq.veldep4.1}) - (\ref{eq.veldep4.4}), we obtain the following system of equations:
\begin{eqnarray}
0 &=& L_{(i_{1}...i_{m}|i_{m+1})} \label{eq.nh11.1} \\
0 &=& M_{,t} -e^{\lambda t} L_{c}Q^{c} \label{eq.nh11.2} \\
0 &=& M_{,i_{1}} + e^{\lambda t} \left( \lambda L_{i_{1}} -2L_{i_{1}i_{2}}Q^{i_{2}} \right), \enskip m>1 \label{eq.nh11.3} \\
0 &=&  L_{(i_{1}...i_{r-1}|i_{r})} +\lambda L_{i_{1}...i_{r}} -(r+1) L_{i_{1}...i_{r}i_{r+1}} Q^{i_{r+1}}, \enskip r=2,3,...,m-1, \enskip m>2 \label{eq.nh11.4} \\
0 &=& M_{,i_{1}} +\lambda e^{\lambda t}L_{i_{1}}, \enskip m=1 \label{eq.nh11.5} \\
0 &=& \lambda L_{i_{1}...i_{m}} +L_{(i_{1}...i_{m-1}|i_{m})}, \enskip m>1 \label{eq.nh11.6}
\end{eqnarray}
where --without loss of generality-- all the non-zero constants $\lambda_{r}$ are fixed to the same non-zero constant $\lambda$. The $m$th-order FI produced from this assumption contains as subcases all the FIs associated with constants $\lambda_{r}$ which are not all the same.

Equation (\ref{eq.nh11.1}) implies that $L_{i_{1}...i_{m}}$ is an $m$th-order generalized KT.

Integrating equation (\ref{eq.nh11.2}), we find
\begin{equation}
M= \frac{e^{\lambda t}}{\lambda}L_{c}Q^{c} +G(q) \label{eq.nh12}
\end{equation}
where $G(q)$ is an arbitrary smooth function.

Replacing (\ref{eq.nh12}) in equations (\ref{eq.nh11.3}) and (\ref{eq.nh11.5}), we find that $G(q)=const\equiv 0$ and the condition:
\begin{equation}
\left( L_{c}Q^{c} \right)_{,i_{1}}= 2\lambda L_{i_{1}i_{2}}(m>1) Q^{i_{2}} -\lambda^{2}L_{i_{1}}. \label{eq.nh13}
\end{equation}

Equation (\ref{eq.nh11.4}) gives the condition
\begin{equation}
L_{(i_{1}...i_{r-1}|i_{r})}= (r+1)L_{i_{1}...i_{r}i_{r+1}} Q^{i_{r+1}} -\lambda L_{i_{1}...i_{r}}, \enskip r=2,3,...,m-1, \enskip m>2 \label{eq.nh14.1}
\end{equation}
while the remaining condition (\ref{eq.nh11.6}) implies that
\begin{equation}
L_{i_{1}...i_{m}}= -\frac{1}{\lambda} L_{(i_{1}...i_{m-1}|i_{m})}, \enskip m>1. \label{eq.nh14.2}
\end{equation}

The associated $m$th-order FI (\ref{FI.5}) is
\begin{equation}
I^{(m)}_{e}= \frac{e^{\lambda t}}{\lambda} \left( \lambda \sum^{m}_{r=1} L_{i_{1}...i_{r}} \dot{q}^{i_{1}} ... \dot{q}^{i_{r}} +L_{c}Q^{c} \right) \label{eq.nh15}
\end{equation}
where $\lambda\neq0$, $L_{i_{1}...i_{m}}(q)$ is an $m$th-order generalized KT satisfying the condition (\ref{eq.nh14.2}), and the remaining totally symmetric tensors $L_{i_{1}...i_{r}}(q)$ with $r=1,2,...,m-1$ and $m>1$ satisfy the conditions (\ref{eq.nh13}) and (\ref{eq.nh14.1}).

The notation $I^{(m)}_{e}$ indicates the $m$th-order FI (upper index) with exponential time-dependence (lower index).
\bigskip

The above complete the proof of Theorem \ref{thm1}.

\section*{Acknowledgments}

We thank Dr Andronikos Paliathanasis for proposing the dynamical system (\ref{eq.aham8a}) - (\ref{eq.aham8b}).

\section*{Data Availability}

The data that supports the findings of this study are available within the article.

\section*{Conflict of interest}

The authors declare no conflict of interest.

\section*{Author contributions}

Conceptualization, Antonios Mitsopoulos and Michael Tsamparlis; Formal analysis, Antonios Mitsopoulos; Methodology, Antonios Mitsopoulos and Michael Tsamparlis; Writing – original draft, Antonios Mitsopoulos; Writing – review and editing, Antonios Mitsopoulos, Michael Tsamparlis and Aniekan Magnus Ukpong.

\end{document}